\documentclass[preprint] {article}
\usepackage{amsmath,amssymb,amsfonts}
\usepackage{geometry}
\newgeometry{vmargin={40mm,27mm}, hmargin={25mm,25mm}}   

\usepackage{graphicx}
\allowdisplaybreaks
\newcommand {\bp}{\begin{pmatrix}}
\newcommand {\ep}{\end{pmatrix}}
\newcommand{\be}{\begin{equation}} \newcommand{\ee}{\end{equation}}
\newcommand{\bea}{\begin{eqnarray}}\newcommand{\eea}{\end{eqnarray}}

\DeclareMathOperator{\sgn}{sgn}

\begin{document}

\title{ Classical Hamiltonian Systems with balanced loss and gain}

\author{Pijush K Ghosh\footnote {{\bf email:} pijushkanti.ghosh@visva-bharati.ac.in}}

\date{ Department of Physics, Siksha-Bhavana, Visva-Bharati University,\\
Santiniketan, PIN 731 235, India.}
\maketitle

\begin{abstract}
Classical Hamiltonian systems with balanced loss and gain are considered in this review.
A generic Hamiltonian formulation for systems with space-dependent balanced loss and gain is discussed.
It is shown that the loss-gain terms may be removed completely through appropriate co-ordinate transformations 
with its effect manifested in modifying the strength of the velocity-mediated coupling.
The effect of the Lorentz interaction in improving the stability of classical solutions as well as
allowing a possibility of defining the corresponding quantum problem consistently on the real line, instead of
within Stokes wedges, is also discussed. Several exactly solvable models based on translational and rotational symmetry 
are discussed which include coupled cubic oscillators, Landau Hamiltonian
etc. The role of ${\cal{PT}}$-symmetry on the existence of periodic solution in systems with balanced loss
and gain is critically analyzed. A few non-${\cal{PT}}$-symmetric Hamiltonian as well as non-Hamiltonian systems
with balanced loss and gain, which include mechanical as well as extended system,  are shown to admit periodic solutions.
An example of Hamiltonian chaos within the framework of a non-${\cal{PT}}$-symmetric system of coupled Duffing oscillator
with balanced loss-gain and/or positional non-conservative forces is discussed. It is conjectured that a
non-${\cal{PT}}$-symmetric system  with balanced loss-gain and without any velocity mediated interaction may admit
periodic solution if the linear part of the equations of motion is necessarily ${\cal{PT}}$ symmetric \textemdash the
nonlinear interaction may or may not be ${\cal{PT}}$-symmetric. Further, systems with velocity mediated interaction need
not be ${\cal{PT}}$-symmetric at all in order to admit periodic solutions. Results related to nonlinear
Schr$\ddot{o}$dinger and Dirac equations with balanced loss and gain are mentioned briefly. A class of solvable models of
oligomers with balanced loss and gain is presented for the first time along with the previously known results.
\end{abstract}

\tableofcontents{}

\section{Introduction}

Dissipation is an ubiquitous phenomenon in nature. There are different approaches to
understand dissipation, instabilities induced by it and controlling its effect in physical
systems\cite{marsden,morrison}. One of the earlier attempts in this
direction was Hamiltonian formulation of system with dissipation. This may seem
counter intuitive, since a Hamiltonian system is conservative,
while dissipation induces the loss of energy to the environment. The apparent conflict
can be resolved if the system plus environment is considered as a larger
system. The Hamiltonian formulation of Bateman oscillator is based on this approach
and describes a system comprising of a damped oscillator plus an  auxiliary system
of an anti-damped oscillator with the same dissipation/anti-dissipation coefficient \textemdash
the dimension of the ambient space is twice that of the target space\cite{bat}.
The flow in the position-velocity state-space preserves the volume, allowing a
Hamiltonian of the system, since the rate of energy-loss of the damped
mode is equal to the rate of gain in energy by the anti-damped oscillator. 
The Bateman oscillator is an example of a system with balanced loss and gain \textemdash the flow
is preserved in the position-velocity state-space although individual degrees of
freedom are subjected to gain/loss.

The Bateman oscillator has been studied extensively over the last ninety years from different
perspectives\cite{morse,bopp,feshback,trikochinsky,dekker-0,rasetti,rabin,jur,bc,palermo,
deguchi}. One of the undesirable features of the Hamiltonian formulation of Bateman oscillator is
the introduction of the auxiliary degree of freedom. With the growing interest and
relevance of ${\cal{PT}}$-symmetric theory\cite{bbook}, the concept of auxiliary system is abandoned
in the interpretation of generalized versions of Bateman oscillator, where damped and anti-damped modes
are coupled and exchange energies. It should be noted that there is no exchange of energies
between the damped and anti-damped modes of the standard Bateman oscillator and a clear distinction between the
two modes exists. Further, there are descriptions of quantum dissipation\cite{caldeira}, where the energy is
deposited from the system to the bath consisting of infinitely many harmonic oscillators so that the energy
can not be transferred back to the system. The sole objective of adding interaction in the generalized models
of  Bateman oscillator is allowing a bi-directional exchange of energies between the damped and the anti-damped modes such that
an equilibrium may be achieved \textemdash the energy is reverted back to the damped oscillator from the
anti-damped oscillator at the same rate as it is deposited to the anti-damped oscillator from the damped oscillator.
The existence of equilibrium state in a ${\cal{PT}}$-symmetric system with balanced loss and gain is worth comparing
to a similar situation arising in the context of non-equilibrium thermodynamics, where the invariance of Schwinger-Keldysh
action under time-reversal symmetry plus time-translation corresponds to thermodynamic equilibrium
\cite{sk}. The Hamiltonian is the generator of the time-translation, while invariance under time-reversal
symmetry conforms to principle of detailed balance for the quantum system. It should be noted here that
${\cal{PT}}$-symmetry may be substituted with a {\it non-standard time-reversal symmetry} within the realm of
quantum mechanics\cite{pkg-ds}. Thus, the conditions on the Schwinger-Keldysh action for thermodynamic equilibrium 
and the equilibrium condition for systems with balanced loss and gain may be identified as similar. 

The basic model of a generalized Bateman oscillator may be described by considering two linearly coupled identical
harmonic oscillators, one of which is subjected to damping and the other to anti-damping with the same strength\cite{cmb}.
This describes an experimentally realized system involving coupled whispering galleries\cite{cmb-1}.
The system consists of two degrees of freedom and there is no concept of auxiliary system \textemdash
the target and the ambient spaces are the same unlike in the case of the standard Bateman oscillator.
The abandoning of  auxiliary degrees of freedom, while considering models of generalized Bateman oscillator,
is indeed a paradigm shift from the traditional treatment of the  standard Bateman oscillator. The coupling
allows an equilibrium state manifested as periodic solutions within some regions in the parameter-space characterizing the
unbroken ${\cal{PT}}$-phase. The existence of equilibrium state is one of the novel features of
a system with balanced loss and gain. The phase-transitions between broken and unbroken ${\cal{PT}}$-phases,
existence of quantum bound states in an unbroken ${\cal{PT}}$-phase, role of exceptional points etc. are
some of the salient aspects of a system with balanced loss and gain\cite{bbook}. 

The central focus of this article is to review known results on classical mechanical systems with balanced loss and gain for
arbitrary number of particles and physically interesting potentials. The criteria for a mechanical system to be identified
as a system with balanced loss and gain is discussed in Sec. \ref{over-balanced loss and gain}. The Hamiltonian formulation
for a generic system with balanced loss and gain is discussed in Sec. \ref{HS} along with a representation of the matrices
appearing in the Hamiltonian. The loss-gain coefficient may be taken to be constant or space-dependent.
It is shown that the loss-gain terms may be removed with its effect manifested in modifying the strength
of the velocity-mediated coupling through appropriate co-ordinate transformations. This result is independent
of any particular representation of the matrices. The effect of the Lorentz interaction in improving the
classical stability as well as defining the quantum problem on the real line is discussed.  In Sec. \ref{sm},
different exactly solvable models based on translational and rotational symmetry are discussed. It is also
shown by constructing $m+1$ integrals of motion for a system with $N=2m$ particles that the system is at least
partially integrable for $N >2$ and completely integrable for $N=2$. The developments in the field
of ${\cal{PT}}$-symmetry played a significant role in interpreting a Hamiltonian of generalized Bateman oscillator
as a system without any auxiliary degrees of freedom. The role of ${\cal{PT}}$-symmetry on the existence of periodic
solution is critically analyzed in Sec. \ref{hpt}. A non-${\cal{PT}}$-symmetric Hamiltonian system of coupled Duffing
oscillators with balanced loss-gain and positional non-conservative forces is shown to admit regular as well as
chaotic dynamics. The periodic solutions along with bifurcation diagrams and Lyapunov exponents are presented.
Further, a non-${\cal{PT}}$-symmetric non-Hamiltonian system of coupled Duffing oscillators with balanced loss
and gain is also shown to admit periodic solution. Other examples of non-${\cal{PT}}$-symmetric Hamiltonian system
with balanced loss and gain which admit periodic solution include Landau Hamiltonian, a dimer model and nonlinear
Schr$\ddot{o}$dinger equation. It is conjectured that a system  with balanced loss-gain and  without any velocity
mediated interaction may admit periodic solution if the linear part of the equations of motion is necessarily
${\cal{PT}}$-symmetric \textemdash the nonlinear interaction may or may not be ${\cal{PT}}$-symmetric. Further,
systems with velocity mediated coupling among different degrees of freedom need not be ${\cal{PT}}$ symmetric at
all in order to admit periodic solutions.
The subject of system with balanced loss and gain has many facets which are not included in this review and a few important
omitted topics are mentioned in Sec. \ref{ot}. The nonlinear Schr$\ddot{o}$dinger and Dirac equations with balanced
loss and gain are reviewed in this section. Further, a class of solvable models of oligomers with balanced loss and
gain is presented for the first time along with the previously known results. Finally, in Sec. \ref{sd}, the results
are summarized along with discussions.

\section{Overviews on Systems with balanced loss and gain}\label{over-balanced loss and gain}

The Bateman oscillator is described by the following set of equations:
\bea
&& \ddot{x} + 2 \gamma \dot{x} + \omega^2 x=0\nonumber \\
&& \ddot{y} - 2 \gamma \dot{y} + \omega^2 y=0
\label{bo}
\eea
\noindent where $\dot{x}=\frac{dx}{dt}$ and $\ddot{x}=\frac{d^2x}{dt^2}$. The $x$ and $y$ degrees of freedom are subjected
to loss and gain, respectively for $\gamma >0$. The Bateman oscillator was originally introduced as a Hamiltonian system
for a dissipative simple harmonic oscillator. The basic idea is to embed the
original system described by the $x$ degree of freedom in a larger system with two degrees of
freedom, where the additional $y$ degree of freedom defines an auxiliary system. The ambient space is defined by
taking both the $x$ and $y$ degrees of freedom together, while $x$ degree of freedom alone describes
the target space. The anti-damped oscillator being the time-reversed version of the damped oscillator and
the vice verse, a Hamiltonian formulation is allowed in the ambient space. The Bateman oscillator has been studied
for almost ninety years from different perspectives with the central theme being
a Hamiltonian description for a dissipative oscillator. A major unwanted feature of Bateman oscillator
is the presence of auxiliary degree of freedom. A few other alternative approaches to study the same
problem with a non-standard Hamiltonian description and without any auxiliary degree of freedom
have also been proposed \textemdash (i) time-dependent
Hamiltonian\cite{bat,ckb}, (ii) a complex Lagrangian\cite{dekker}, and (iii) different Hamiltonians
for different parameter regimes by using modified Prelle-Singer method\cite{ml}. All these methods have
their own merits and demerits with a common goal \textemdash Hamiltonian description of a dissipative system.
However, abandoning the idea of auxiliary system for the case of Bateman oscillator
and reinterpreting it as conservative system with balanced loss and gain for which the ambient and the target spaces
are the same with two degrees of freedom has not been explored until recently.

The Bateman oscillator is one of the simplest examples of a system
with balanced loss and gain. The flow in the phase space preserves its volume due to the balanced loss and gain
\textemdash any imbalance leads to either growing or decaying volume. The flow in the phase-space
preserves its volume even if interaction terms are added by coupling the two degrees of freedom:
\bea
&& \ddot{x} + 2 \gamma \dot{x} + \omega^2 x + G_1(x,y)=0\nonumber \\
&& \ddot{y} - 2 \gamma \dot{y} + \omega^2 y + G_2(x,y)=0,
\label{bo2}
\eea
\noindent where $G_1(x,y)$ and $G_2(x,y)$ are arbitrary functions. A different viewpoint may be considered
where the mechanical model described by Eq. (\ref{bo2}) with two degrees of freedom is taken as a whole
system without any auxiliary degree of freedom. In other words, the target and the ambient spaces are
taken to be the same. This is a paradigm shift in the interpretation of models with balanced loss and gain
and leads to a new type of interacting system with important physical consequences.
The modified system defined by Eq. (\ref{bo2}) may admit periodic solutions for specific choices of
$G_{1,2}(x,y)$ which is not possible for the Bateman oscillator, i.e. $G_1(x,y)=G_2(x,y)=0$.  For example, Eq. (\ref{bo2})
with $G_1=\epsilon y, G_2=\epsilon x, \epsilon \in \Re$, represents a mathematical model\cite{cmb} for an experimentally
realized ${\cal{PT}}$ symmetric coupled resonators and admits periodic solutions within suitable ranges of
the parameters\cite{cmb-1}. Similarly, a two particle Calogero-type model with balanced loss and gain admitting periodic
solutions may be reproduced for $G_1=\alpha (x-y)^{-3}, G_2=-\alpha (x-y)^{-3}, \alpha \in \Re$\cite{sg-calo}. There
are many examples of systems with balanced loss and gain admitting periodic solutions in certain regions
of the parameters space\cite{bend-2,igor,khare,pkg-ds,sg-calo-cor,ds-pkg,pkg,pkg-pr}.

The concept of a system with balanced loss and gain may be generalized for arbitrary $N$ degrees of freedom.
The system is governed by the equations of motion,
\bea
\ddot{X} -2 {\cal{D}} \dot{X} + G(x_1, x_2, \dots, x_N)=0,
\label{bo3}
\eea
\noindent where the co-ordinates of the $N$ particles are denoted as $x_1, x_2, \dots, x_N$ which are
elements of the column matrix $X$. In particular, $X^T \equiv (x_1, x_2, \dots, x_N)$, where $X^T$ denotes the
transpose of $X$. The  term linear in $\dot{X}$ contains information on the nature of
loss-gain as well as velocity mediated coupling among the particles depending on the explicit form
of the $N \times N$ matrix ${\cal{D}}$. In general, ${\cal{D}}$ may be decomposed as,
\bea
{\cal{D}} = D +D_{SO} + {\cal{D}}_A
\eea
\noindent where $D$ is a diagonal matrix, $D_{SO}$ is a symmetric matrix with vanishing diagonal elements
and ${\cal{D}}_A$ is an anti-symmetric matrix. The symmetric matrix ${\cal{D}}_S=D +D_{SO}$ is decomposed
in terms of its
diagonal and off-diagonal parts, $D$ and $D_{SO}$, respectively. The Lorentz interaction is described by
${\cal{D}}_A$, while velocity mediated non-Lorentzian interaction is encoded in $D_{SO}$. The Lorentz interaction
appears in the description of many physical systems\cite{ll}, while velocity mediated non-Lorentzian
interaction appears in the synchronization of different types of oscillators\cite{nli}.
The loss and gain in
the system are introduced via the matrix $D$. In general, the matrix ${\cal{D}}$ may be space-dependent and
thereby, allowing space-dependent loss-gain terms in the system.
The space-mediated coupling among different degrees of freedom is encoded through the $N$-component
field $G(x_1,x_2,\dots,x_N)$ with its components $G_i \equiv G_i(x_1, x_2, \dots, x_N), i=1, 2, \dots, N$.

The criteria for the system described by Eq. (\ref{bo3}) to be non-dissipative may be determined by studying
the time-evolution of the flow in the $2N$ dimensional position-velocity state space. The flow preserves the
volume in the position-velocity state-space for a non-dissipative system, while the volume either grows or decays
depending on whether the system is anti-damped or damped, respectively. Introducing a $2N$-component vector
$\xi$ and a field $\eta$,
\bea
\xi \equiv \bp X\\ \dot{X} \ep, 
\eta_i \equiv \xi_{N+i}, \
\eta_{N+i} = - G_i(\xi_i, \dots, \xi_N) + \sum_{k=1}^N {\cal{D}}_{ik}(\xi_i, \dots, \xi_N) \xi_{N+k}
\eea
\noindent Eq. (\ref{bo3}) may be rewritten in terms of $2N$ first-order coupled differential equations as
$\dot{\xi}= \eta$. The flow preserves the volume $V=\prod_{i=1}^{2N} d \xi_i$ in the position velocity state
space provided $\eta$ is solenoidal, i. e.
\bea
\sum_{i=1}^{2N} \frac{\partial \eta_i}{\partial \xi_i}=0 \ \ \Rightarrow Tr(D)=0,
\eea
\noindent where $Tr$ denotes trace of a matrix. The two dimensional examples in Eqs. (\ref{bo}) and (\ref{bo2}),
when cast into the form of Eq. (\ref{bo3}), contain ${\cal{D}}=D= \gamma \ diag(-1,1)$ which is indeed traceless. For a system
with $N>2$ degrees of freedom, the traceless condition may be implemented in several ways depending on the actual
physical scenario. The vanishing trace of $D$ is the criteria for a model governed by Eq. (\ref{bo3}) to be identified
as a system with balanced loss and gain for which the flow in the position-velocity state space preserves the volume
in spite of the fact that individual degrees of freedom are subjected to loss or gain.

\section{Hamiltonian system}\label{HS}

The Hamiltonian formulation of a system with balanced loss and gain is a non-trivial problem. For example,
the Hamiltonian of two linearly coupled identical Duffing oscillators, one of which
is subjected to gain and the other with an equal amount of loss is unknown. The same
system with additional nonlinear interaction may be shown to be Hamiltonian\cite{bkt}.
It should be mentioned here that there are well known methods to construct Lagrangian
and Hamiltonian from a given set of equations of motion. The inverse variational problem
was initiated by Helmholtz\cite{bkt}. However, a successful implementation of the scheme
becomes nontrivial for many-particle systems with nonlinear interaction. Further, a system
may not admit a Lagrangian-Hamiltonian formulation at all\cite{santili}. Within this background,
the Hamiltonian formulation for a class of systems with balanced loss and gain is described below.

The Hamiltonian of the Bateman oscillator has the expression,
\bea
&& H_B=P_x P_y + \gamma \left (y P_y-x P_x \right ) + \left (\omega^2-\gamma^2 \right ) xy\nonumber \\
&& P_x=\dot{y}-\gamma y, \ P_y=\dot{x} + \gamma x,
\eea
\noindent where the canonical conjugate momenta corresponding to $x$ and
$y$ are denoted as $P_x$ and $P_y$, respectively. The Hamilton's equation of motion reproduces
Eq. (\ref{bo}). The Hamiltonian $H_B$ can be rewritten in terms of the generalized momenta
$\Pi_x$ and $\Pi_y$ as,
\bea
H_B=\Pi_x \Pi_y + \omega^2 x y, \ \ \Pi_x=P_x + \gamma y, \Pi_y=P_y - \gamma x.
\eea
\noindent The following points may be noted:
\begin{itemize}
\item The expressions for $\Pi_{x,y}$ are similar to that of the generalized momenta for a particle
in an uniform external magnetic field with magnitude $\gamma$ and along the direction perpendicular
to the $x-y$ plane. There is no magnetic field in the system. So, $A_x:=\gamma y, A_y:=-\gamma x$
may be interpreted as `fictitious gauge potentials' leading to the uniform `fictitious magnetic field'
with magnitude $\gamma$.
\item The quadratic term involving the generalized momenta is not positive-definite,
$\Pi_x \Pi_y=\Pi_+^2 - \Pi_-^2, \ \Pi_{\pm}= \frac{1}{2} ( \Pi_x \pm \Pi_y )$ and the system
may be interpreted as defined in the background of a pseudo-Euclidean metric with signature
$(1,-1)$.
\end{itemize}
\noindent These two features are present for known Hamiltonian systems with balanced loss and gain
and will be used as essential inputs for constructing Hamiltonian for a generic many-particle
system with balanced loss and gain.

The Hamiltonian for a general system with balanced loss and gain is taken to be of the form,
\bea
H=\Pi^T {\cal{M}} \Pi + V(x_1, x_2, \dots, x_N),
\label{hami}
\eea
\noindent where ${\cal{M}}$ is an $N \times N$ real symmetric matrix and $\Pi=(\pi_1, \pi_2, \dots,
\pi_N)^T$ denotes $N$ component generalized momenta. The matrix ${\cal{M}}$ is non-singular so that
${\cal{M}}^{-1}$ exists and is not necessarily semi-positive definite. The matrix ${\cal{M}}$ may be
interpreted as a constant background metric as in the case of Bateman oscillator. A semi-positive definite ${\cal{M}}$,
if exists, may be interpreted either as a metric or a mass-matrix. The generalized momenta $\Pi$ has
the expression,
\bea
\Pi=P+A F(X),
\label{gm}
\eea
\noindent where $P=(p_1, p_2, \dots, p_N)^T$  is the conjugate momentum
corresponding to the coordinate$X$, $F(X)=(F_1, F_2, \dots F_N)^T$ is $N$
dimensional column matrix whose entries are functions of coordinates and $A$ is an $N \times N$
anti-symmetric matrix. It may be noted that $a_i=\sum_{k=1}^N A_{ik} F_k$ can be interpreted as
gauge potentials. In general, $a_i$ may be decomposed as $a_i=a_i^R+a_i^F$ in terms of realistic
gauge potentials $a_i^R$ and fictitious gauge potentials $a_i^F$. The realistic gauge potential
describes external magnetic field in the system, while the fictitious gauge potential leads to
loss-gain terms as in the case of Bateman oscillator. With suitable choices of $F$, space-dependent gain-loss terms
are allowed in the system. The Hamiltonian $H$ in Eq. (\ref{hami}) reduces to the Hamiltonian of Ref. \cite{cmb}
describing coupled harmonic oscillators with balanced loss-gain for the following choices of ${\cal{M}}, A, F, V$
and identification $x_1=x, x_2=y, p_1=P_x, p_2=P_y$:
\bea
{\cal{M}}=\frac{1}{2} \sigma_1, \ A= i \gamma \sigma_2, \ F_1=x, \ F_2=y, \ V(x,y)=\omega^2 x y +\frac{\beta}{2}
\left (x^2 + y^2 \right )
\eea where $\sigma_i, i=1, 2, 3$ are the Pauli matrices. The Hamiltonian $H_B$ of the Bateman oscillator is
reproduced for $\beta=0$. The Lagrangian corresponding to the Hamiltonian $H$ in Eq. (\ref{hami}) may be
derived as follows:
\bea
{\cal{L}} = \frac{1}{4}\dot{X}^T {\cal{M}}^{-1} \dot{X} -\frac{1}{2} (\dot{X}^T
AF +F^T A^T \dot{X}) - V(x_1, x_2, \dots, x_N).
\label{lag}
\eea
\noindent The kinetic energy term in the Lagrangian is not necessarily positive-definite. The presence of
terms linear in velocity is essential for incorporating loss-gain in the system. These are general characteristics
of Bateman oscillator and its generalized versions.

The equations of motion following from the Hamiltonian (\ref{hami}) or the Lagrangian (\ref{lag}) reads,
\bea
\ddot{X}-2{\cal{M}}R\dot{X}+2{\cal{M}}\frac{\partial V}{\partial X}=0,
\label{X}
\eea
\noindent where the anti-symmetric matrix $R$, the Jacobian $J$ and
$\frac{\partial V}{\partial X}$ are defined as follows:
\be
R \equiv AJ-(AJ)^T, \ \
[J]_{ij} \equiv \frac{\partial F_i}{\partial x_j},\ \
\frac{\partial V}{\partial X}\equiv
\left(\frac{\partial V}{\partial x_i}, \frac{\partial V}{\partial x_2},
\dots \frac{\partial V}{\partial x_N}\right)^T.
\label{X-supp}
\ee
\noindent The matrix $R$ is anti-symmetric by definition, while $J$ has no specified symmetry.
However, $J$ can be made to be symmetric or anti-symmetric with suitable choices of the
$N$-component field $F$. The equations of motion (\ref{X}) reduces to Eq. (\ref{bo3}) with the following
identifications:
\be
{\cal{D}}={\cal{M}} R, \ \ G = 2 {\cal{M}} \frac{\partial V}{\partial X}.
\label{consi}
\ee
\noindent The criteria $Tr({\cal{D}})=0$ for a balanced loss and gain system is automatically satisfied, since
${\cal{M}}$ is symmetric and $R$ is anti-symmetric. The construction of Hamiltonian for a given Eq.
(\ref{bo3}) now reduces to finding solutions for the equations in Eq. (\ref{consi}). It is
known\cite{rodman} that a matrix ${\cal{D}}$ that is similar to $-{\cal{D}}$ can always be decomposed as a
product of a symmetric and an anti-symmetric matrices.
However, finding expressions for $A$ and $J$ from a known $R$ is a non-trivial problem, particularly
for the case of space-dependent loss-gain terms.
Further, the problem of finding $V$ for a given $G$ involves solving $N$ coupled first order differential
equations $\frac{\partial V}{\partial X}=\frac{1}{2} {\cal{M}}^{-1} G$ which, in general, may elude a closed
form expression for $V$. Nevertheless, the Hamiltonian of a large number of physical system with balanced loss and gain
may be constructed via specific representation of matrices ${\cal{M}}$, $R$ and ${\cal{D}}$ and closed
form expression for $V$\cite{pkg-ds,ds-pkg,pkg}.

The Hamiltonian for systems with space-independent loss and gain terms for $N=2$ has been considered earlier\cite{bat,cmb}.
The general formalism described above may be used to construct Hamiltonian system with space-dependent loss-gain terms.
For example, a coupled Van der Pol-Duffing oscillator model with balanced loss and gain may be obtained by choosing,
\bea
{\cal{M}}=\sigma_1 + \alpha^2 I_2, \ A=-\frac{i \gamma}{2} \sigma_2, \ F_i = a_i x_i + b_i x_i^3, \ 
V(x_1,x_2)= \frac{\omega^2}{2} x_1 x_2 + \frac{\beta}{4} \left (x_1^2 + x_2^2\right ) + g x_1^3 x_2,
\label{vdpd}
\eea
\noindent where $I_2$ is the $2 \times 2$ identity matrix and $\omega, \alpha, \beta, \gamma, a_i, b_i, g$ are real
constants. The choice of the matrix $A$ and the field $F(x_1,x_2)$ uniquely fixes the generalized momenta $\Pi$, while
that of the matrix ${\cal{M}}$ and  the potential $V(x_1,x_2)$ completely specifies the Hamiltonian. The expressions
for the matrices $R(x_1,x_2)$ and ${\cal{D}}(x_1,x_2)$ can be computed by using Eqs. (\ref{X-supp}) and (\ref{consi}),
respectively,
\bea
R=-\frac{i \gamma}{2} \sigma_2 Q(x_1,x_2), \ {\cal{D}}=\frac{\gamma}{2} Q(x_1, x_2) \left ( \sigma_3 -
i \alpha^2 \sigma_2 \right ), \ Q(x_1,x_2)\equiv a_1 +a_2 + 3 \left (b_1 x_1^2 + b_2 x_2^2 \right )
\label{vdpd-1}
\eea
\noindent Eq. (\ref{lag}) describes a coupled Van der Pol-Duffing oscillator\cite{pr-pkg}, 
\bea
&& \ddot{x}_1 - \gamma Q(x_1,x_2) \left ( \dot{x}_1 - \alpha^2 \dot{x}_2 \right ) + \omega^2 x_1 + \beta x_2 +
g x_1^3=0\nonumber \\
&& \ddot{x}_2 + \gamma Q(x_1,x_2) \left ( \dot{x}_2 - \alpha^2 \dot{x}_1 \right ) +
\omega^2 x_2 + \beta x_1 + 3 g x_1^2 x_2=0
\label{vdpd-2}
\eea
\noindent where $\gamma Q$ is the space-dependent loss-gain coefficient, $\gamma \alpha^2 Q$ is the
external magnetic field, $\omega$ is the angular frequency of the harmonic oscillator. The strengths of the space-mediated
linear and non-linear couplings are $\beta$ and $g$, respectively. The above equation may be reduced to a few known models
with appropriate choices of the parameters. For example,
the Hamiltonian of coupled oscillators with constant balanced loss and gain\cite{cmb} is obtained for
$a_1=a_2=1, b_1=b_2=\alpha=g=0$. The Bateman oscillator is obtained if the linear coupling is switched off additionally
by taking $\beta=0$. The Hamiltonian formulation of the above system for $\beta=\alpha=g=0$, $a_1=a_2=\frac{1}{2}$
and $b_1 \rightarrow -\frac{b_1}{3}, b_2 \rightarrow -\frac{b_2}{3}$ has been considered earlier\cite{sagar} which describes
coupled Van der Pol oscillators. The $x_1$ degree of freedom describes a Van der Pol-Duffing oscillator for
$b_2=\beta=\alpha=0, a_1=a_2=\frac{1}{2}, b_1 \rightarrow -\frac{b_1}{3}$ and the $x_2$ degree of freedom is
unidirectionally coupled to it. This also provides a Hamiltonian formulation for Van der Pol-Duffing oscillator in an
ambient space of two dimensions, much akin to the case of Bateman oscillator. The Hamiltonian of Ref. \cite{pkg-pr}
is obtained for  $a_1=a_2=1, b_1=b_2=\alpha=0$. The system has rich dynamical properties for the generic values of
the parameters\cite{pr-pkg}.

Generalizations to $N=3$ may be achieved in several ways depending on the particular physical contexts. For example,
the non-vanishing elements of the $3 \times 3$ matrices ${\cal{M}}$ and $A$ may be chosen as
${\cal{M}}_{12}={\cal{M}}_{21}={\cal{M}}_{33}=1$, ${\cal{M}}_{11}={\cal{M}}_{22}=\alpha^2$,
$A_{12}=-A_{21}=-\frac{\gamma}{2}$ along with  $F_3=0$ and $F_1, F_2$ as given in Eq. (\ref{vdpd}). The $x_3$ degree of freedom is neither subjected to gain/loss nor
it is coupled to $x_1$ and $x_2$ degrees of freedom via velocity mediated coupling, since all elements of ${\cal{D}}$ are zero except for $D_{11}=-D_{22}=\frac{\gamma}{2} Q(x_1,x_2)$.
The coupling among $x_1, x_2$ and $x_3$ degrees of freedom can be incorporated via the potential $V\equiv V(x_1,x_2,x_3)$. A particular choice of $V$ and
the resulting equation of motions are described below,
\bea
&& V(x_1,x_2,x_3)= \frac{\omega^2}{4} \left ( 2 x_1 x_2 + x_3^2 \right ) + \frac{\beta}{4} \left (x_1^2 + x_2^2 + 2 x_1 x_3 + 2 x_2 x_3 \right ) +
\alpha x_1^3 x_2 + \frac{\delta}{8} x_3^4 \nonumber \\
&& \ddot{x}_1 - \gamma Q(x_1,x_2) \left ( \dot{x}_1 - \alpha^2 \dot{x}_2 \right ) +
\omega^2 x_1 + \beta \left ( x_2 + x_3 \right ) + \alpha x_1^3=0\nonumber \\
&& \ddot{x}_2 + \gamma Q(x_1,x_2) \left ( \dot{x}_2 -\alpha^2 \dot{x}_1 \right ) +
\omega^2 x_2 + \beta \left ( x_1 + x_3 \right ) + 3 \alpha x_1^2 x_2=0\nonumber \\
&& \ddot{x}_3 + \omega^2 x_3 + \beta \left (x_1 + x_2 \right ) + \delta x_3^3=0
\eea
\noindent where the undamped Duffing oscillator described by $x_3$ degree of freedom linearly couples with the $x_1, x_2$ degrees of freedom.
Generalizations to $ N > 3$ may be continued in a similar way and is not discussed. A particular representation of matrices for arbitrary
$N$ is presented in Sec. \ref{header-mat} and several examples are discussed in Refs. \cite{pkg-ds,sg-calo,sg-calo-cor,ds-pkg,pkg,pkg-pr}.

\subsection{Hiding the loss-gain terms}

The Bateman oscillator can be rewritten as,
\bea
\ddot{z}_+ + 2 \gamma \dot{z}_- + \omega^2 z_+=0,
\ddot{z}_- + 2 \gamma \dot{z}_+ + \omega^2 z_-=0,\ \
\label{map}
\eea
\noindent in a rotated co-ordinate system $z_{\pm} = \frac{1}{\sqrt{2}} ( x \pm
y ) $. The loss-gain terms are absent in Eq. (\ref{map}), since the equation of motion
for $z_+$ does not contain a term $\dot{z}_+$ and similarly the equation of motion
for $z_-$ does not contain a term $\dot{z}_-$. Further, the system may be
interpreted as  defined in the background of a pseudo-Euclidean metric with signature $(1,-1)$ and the
particle is subjected to external magnetic field proportional to $\gamma$. This is a generic feature
of Hamiltonian systems with balanced loss and gain. The loss-gain terms may
always be hidden in a specific coordinate system. The real symmetric matrix
${\cal{M}}$ can be diagonalized by an orthogonal matrix $\hat{O}$, i.e.
$M_D= \hat{O}^T {\cal{M}} \hat{O}$. Defining a rotated co-ordinate system 
$\tilde{X} = \hat{O}^T X$ and $\tilde{P} = \hat{O}^T P$, eq. (\ref{bo3}) can be rewritten as,
\bea
\ddot{\tilde{X}} - 2 \left ( M_D \tilde{R} \right ) \dot{\tilde{X}} +
2 M_D \frac{\partial V}{\partial \tilde{X}}=0, \ \tilde{R} = O^T R O
\eea
\noindent The matrix $\tilde{R}$ is anti-symmetric and each diagonal element of
the matrix $M_D \tilde{R}$ is zero, i.e. $[M_D \tilde{R}]_{ii}=0 \ \forall \ i$.
The loss-gain terms are hidden in the co-ordinate system $\tilde{X}$ with the
effect manifested by modifying the velocity mediated coupling terms. The general
formalism may be exemplified with the Van der Pol-Duffing oscillator model described
by Eqs. (\ref{vdpd},\ref{vdpd-1},\ref{vdpd-2}) for which the matrices $\hat{O}, M_D$ and $\tilde{R}$
have the expressions,
\bea
\hat{O}=\frac{1}{\sqrt{2}} \left ( \sigma_1 + \sigma_3 \right ), \ M_D=\alpha^2 I_2 + \sigma_3,
\ \tilde{R}=\frac{i \gamma}{2} Q \sigma_2, \ \Rightarrow
M_D \tilde{R}=\frac{\gamma Q}{2} \left ( \sigma_1 + i \alpha^2 \sigma_2 \right )
\eea
\noindent It is clearly seen that the diagonal elements of the matrix $M_D \tilde{R}$ are zero,
i.e. $[M_D \tilde{R}]_{11}= [M_D \tilde{R}]_{22}=0$. Further, the off-diagonal elements are 
$[M_D \tilde{R}]_{12}= \frac{\gamma Q}{2} \left ( 1 +\alpha^2 \right ), \ 
[M_D \tilde{R}]_{21}= \frac{\gamma Q}{2} \left ( 1-\alpha^2 \right )$ which can not be interpreted
as describing pure Lorentz force, since $[M_D \tilde{R}]_{12} \neq - [M_D \tilde{R}]_{21}$.

The system may be defined in the background of a pseudo-Euclidean metric through
a canonical scale transformation.  The matrix $M_D$ and two other matrices $S$ and $\eta^a$ are defined
as follows:
\be
[M_D]_{ij}= \delta_{ij} \sgn(\lambda_i) {\lvert \lambda_i \rvert}, \ \
[S]_{ij}=\delta_{ij} \sqrt{{\lvert \lambda_i \rvert}}, \ \
[\eta^{(a)}]_{ij}= \delta_{ij} \sgn(\lambda_i), \ \ a=1, 2, \dots, N+1
\label{eta}
\ee
\noindent where $\lambda_i$'s are the eigenvalues of the matrix ${\cal{M}}$
and $\sgn(x)$ is the signum function. The parameter space of the system may
be characterized in terms of at most $N+1$ distinct regions depending on the
number of negative eigenvalues of the matrix ${\cal{M}}$. The superscript $a$ in
$\eta^{(a)}$
identifies the Region-$a$ in the parameter space corresponding to $a-1$ negative
eigenvalues of ${\cal{M}}$. The matrix $\eta^{(a)}$ is to be interpreted
as the background metric for an effective description of the system defined by
the Hamiltonian $H$ and equations of motion following from it in Eqs.
({\ref{hami}}) and ({\ref{X}}), respectively. The canonical scale transformation is defined as follows:
\be
{\cal{X}}= S^{-1} \tilde{X}, \ \ {\cal{P}}= S \tilde{P}, \ \
\label{trans-cor}
\ee
\noindent where ${\cal{X}}\equiv ({\cal{X}}_1, {\cal{X}}_2, \dots,
{\cal{X}}_N)^T$ and ${\cal{P}}\equiv ({\cal{P}}_1, {\cal{P}}_2, \dots,
{\cal{P}}_N)^T$. The purpose of the scale transformation is to normalize
the eigenvalues of $M$ to $\pm 1$ i.e.
$S^{-1} \hat{O}^T {\cal{M}} \hat{O} S^{-1}=\eta^{(a)}$.
The Hamiltonian and the equations of motion
resulting from the transformation in different regions have the forms:
\bea
&&H^{(a)}={\hat{\Pi}}^T \eta^{(a)} {\hat{\Pi}} + {\cal{V}}({\cal{X}}_1, {\cal{X}}_2,
\dots, {\cal{X}}_N),\nonumber \\
&& \ddot{\cal{X}} -  2 \eta^{(a)} {\cal{R}} \dot{\cal{X}} +
2 \eta^{(a)} \left ( \frac{\partial {\cal{V}}}{\partial {\cal{X}}} \right )=0,
\ \ {\cal{V}}({\cal{X}}_1, {\cal{X}}_2, \dots, {\cal{X}}_N)= V(x_1, x_2,
\dots, x_N),
\label{hami-eq}
\eea
\noindent where ${\cal{R}}=S \tilde{R} S$ and the transformed generalized momenta $\Pi$ after rotation and
the scale transformation is denoted as,
$\hat{\Pi} \equiv S\hat{O}^T \Pi= {\cal{P}} + \frac{\cal{R}}{2} {\cal{X}}$.
The quadratic term in momenta for Hamiltonians $H^{(1)}$ and $- H^{(N+1)}$ are
semi-positive definite, while $H^{(a)}, 2 \leq a \leq N$ are not definite.
The loss-gain terms of Eq. (\ref{X}) are absent in Eq. (\ref{hami-eq}). It
may be noted that Eq. (\ref{hami-eq}) contains velocity mediated Lorentzian and
non-Lorentzian interaction in Region-$a, 2 \leq a \leq N$, while only Lorentzian
interaction is present in Region-1 and Region-N+1.

The formalism for an effective description of the system in the  background of a pseudo-Euclidean metric may
be exemplified with the Van der Pol-Duffing oscillator model described
by Eqs. (\ref{vdpd},\ref{vdpd-1},\ref{vdpd-2}). The matrix ${\cal{M}}$ has eigenvalues $\lambda_1 = \alpha^2 +1$
and $\lambda_2 = \alpha^2 - 1$. The parameter space may be divided into two regions: (i) Region-I: $\alpha^2 > 1$
with an Euclidean metric $\eta^{(1)}=I_2$, (ii)  Region-II: $\alpha^2 < 1$ with a pseudo-Euclidean metric
$\eta^{(2)}= \sigma_3$. The new co-ordinates ${\cal{X}}$, the matrix ${\cal{R}}$, the space-dependent co-efficient
${\cal{Q}}({\cal{X}}_1, {\cal{X}}_2)$ and the potential ${\cal{V}}({\cal{X}}_1, {\cal{X}}_2)$ have the expressions:
\bea
&& {\cal{X}}_1=\frac{1}{\sqrt{2 {\vert \lambda_1 \vert}}} \left ( x_1 + x_2 \right ), \
{\cal{X}}_2=\frac{1}{\sqrt{2 {\vert \lambda_2 \vert}}} \left ( x_1 - x_2 \right ), \
{\cal{R}}=\frac{i \gamma}{2} Q \lambda \sigma_2\nonumber \\
&& {\cal{Q}}({\cal{X}}_1,{\cal{X}}_2)= a_1 + a_2 + \frac{3(b_1+b_2)}{2} \left ( {\vert \lambda_1 \vert} {\cal{X}}_1^2 + 
{\vert \lambda_2 \vert} {\cal{X}}_2^2 \right ) + 3 \left ( b_1 - b_2 \right ) \lambda {\cal{X}}_1 {\cal{X}}_2\nonumber \\
&& {\cal{V}}({\cal{X}}_1,{\cal{X}}_2)= \frac{\Omega_+}{4} {\cal{X}}_1^2 + \frac{g \lambda_1^2}{4} {\cal{X}}_1^4
-\frac{\Omega_-}{4} {\cal{X}}_2^2 - \frac{g \lambda_2^2}{4} {\cal{X}}_2^4
+ \frac{g \lambda}{2} {\cal{X}}_1 {\cal{X}}_2 \left ( {\vert \lambda_1 \vert} {\cal{X}}_1^2 -
{\vert \lambda_2 \vert} {\cal{X}}_2^2 \right )
\eea
\noindent where $\lambda \equiv \sqrt{{\vert \lambda_1 \vert} {\vert \lambda_2 \vert}}$,
$\Omega_+ \equiv \left ( {\omega^2+\beta} \right ) {\vert \lambda_1 \vert}$ and
$\Omega_- \equiv \left ( {\omega^2-\beta} \right ) {\vert \lambda_2 \vert}$.
\begin{itemize}
\item Region-I: The matrix ${\cal{M}}$ is positive-definite. The effective description of the system is in the background
of a two dimensional Euclidean metric. The particle is subjected to Lorentz interaction and there is no other
velocity-mediated interaction.
\item Region-II: The matrix ${\cal{M}}$ is indefinite. The effective description of the system is in the background
of a two dimensional pseudo-Euclidean metric  with the signature $(1-1)$. The particle is subjected to non-Lorentzian
velocity-mediated interaction.
\end{itemize}
Another simple example with two degrees
of freedom is given in Sec. \ref{header-li} which explains the general idea presented above.

\subsection{Representation of matrices}\label{header-mat}

Several representations of the matrices for a vanishing Lorentz interaction and constant loss-gain terms are
presented in Ref. \cite{pkg-ds}.  A few representations for space-dependent loss-gain terms are included in
Refs. \cite{ds-pkg} and \cite{pkg} for vanishing and non-vanishing Lorentz interaction, respectively.
A particular representation for $N = 2m, m \in \mathbb{Z^{\geq}}$ with pair-wise balancing of space-dependent
loss-gain terms is discussed in this article. The matrix ${\cal{M}}$ is chosen as,
\be
{\cal{M}} = M + \alpha^2 I_{2m}, \alpha \in \mathbb{R},
\label{mnew}
\ee
\noindent where $M$ is a traceless $2m \times 2m$ symmetric matrix that anti-commutes with $R$, i.e. $\{M,R\}=0$
and $I_{2m}$ is the $2m \times 2m$ identity matrix. The substitution of ${\cal{M}}$ in eq. (\ref{mnew})
to the expression ${\cal{D}}={\cal{M}} R$ gives,
\be
{\cal{D}} = \underbrace{MR}_{{\cal{D}}_S} + \underbrace{\alpha^2 R}_{{\cal{D}}_A}
\ee
\noindent The relations $M^T=M, R^T=-R, \{M,R\}=0$ ensure that ${{\cal{D}}_S}=MR$ is a symmetric matrix.
The parameter $\alpha$ controls the strength of
the Lorentz interaction in the system and $\alpha=0$ corresponds to vanishing Lorentz interaction.

The $m$ functions $Q_i \equiv Q_i(x_{2i-1}, x_{2i})$ are introduced as,
\be
Q_a(x_{2a-1},x_{2a}) = Tr(V_a^{(2)}), \ \
V_a^{(2)} \equiv
\bp
{\frac{\partial F_{2a-1}}{\partial x_{2a-1}}}
& {\frac{\partial F_{2a-1}}{\partial x_{2a}}}\\
{\frac{\partial F_{2a}}{\partial x_{2a-1}}} &
{\frac{\partial F_{2a}}{\partial x_{2a}}}
\ep.
\label{repq}
\ee
\noindent The representation of the matrices is specified as follows:
\bea
M=I_m\otimes\sigma_1,\ \ \  \ A=\frac{-i\gamma}{2}I_m\otimes\sigma_2,\ \
{\cal{D}}_S=\gamma \chi_m\otimes \sigma_3, \ \ [\chi_m]_{ij}=\frac{1}{2}\delta_{ij}
Q_i(x_{2i-1}, x_{2i}),
\label{repre}
\eea
\noindent where $I_m$ is $m \times m$ identity matrix. The matrix $R$ may be determined
by noting that the matrix $J$ has a block-diagonal form for the choices of $F_i\equiv F_i(x_{2i-1},x_{2i})$:
\be
R = \frac{\gamma}{2} \sum_{i=1}^m U_i^{(m)} \otimes
\bp
0 && - Q_i(x_{2i-1},x_{2i})\\
Q_i(x_{2i-1},x_{2i}) && 0
\ep
,
\ \ [U_a^{(m)}]_{ij} = \delta_{ia} \delta_{ja}.
\label{repare}
\ee
\noindent This completely specifies the representation of the matrices for pair-wise balancing of loss-gain terms
and vanishing non-Lorentzian velocity mediated coupling, i.e. ${\cal{D}}_S=D, D_{SO}=0$. A representation for the
case $D_{SO} \neq 0$ may be found in Ref. \cite{pkg}.

\subsubsection{Effect of Lorentz interaction:}\label{header-li}

The Lorentz interaction in the system vanishes for $\alpha=0$ for which ${\cal{M}}=M$ and ${\cal{D}}={\cal{D}}_S$.
It is known\cite{pkg-ds} that the matrices $M, R$ and ${\cal{D}}_S$ anti-commute with each other and, hence,
\bea
\{{\cal{M}},R\}=0, \ \{{\cal{M}},{\cal{D}}_S\}=0, \{{\cal{D}}_S,R\}=0.
\eea
\noindent An immediate consequence is that the matrix ${\cal{M}}=M$ is indefinite and  corresponding to each of its $m$
positive eigenvalues $\lambda_i$, there exists an eigenvalue $- \lambda_i$.
The term $\Pi^T {\cal{M}} \Pi$ in the Hamiltonian is not semi-positive definite and this has important consequences for
the classical as well as quantum systems.
\begin{itemize}
\item The Hamiltonian is not bounded from below even for a $V$ with a well-defined lower bound. Consequently, the 
question of stability of the system is much more involved compared to separable Hamiltonian. In general, the
Lagrange-Dirichlet theorem\cite{dirichlet,marsden} for Hamiltonian system does not give any conclusive results on the
stability of the solutions.

\item The quantum problem is not well-defined on the real line and a well-defined ground state
does not exist. However, if the quantum Hamiltonian is defined in appropriate Stokes wedges, the system may admit
well defined bound states\cite{pkg-ds,cmb,sg-calo,sg-calo-cor,ds-pkg,pkg-pr}. This is consistent from the
viewpoint of axiomatic foundations of quantum mechanics. However, no experimentally realizable system of this
type has been found so far.
\end{itemize} The Hamiltonian of the Bateman oscillator is unbounded from below. However, for specific choices of $V$, the Hamiltonian
may be bounded from below and admit periodic solutions. The Hamiltonian of Ref. \cite{cmb} and examples
considered in Refs. \cite{pkg-ds,sg-calo,sg-calo-cor,ds-pkg,pkg,pkg-pr} are
bounded from below/above within some regions in the parameter-space. These examples admit stable periodic
solutions at the classical level and quantum bound states.

The Lorentz interaction is switched on for $\alpha \neq 0$ and introduces external magnetic field in the
system. On the other hand, the loss-gain terms may be interpreted as `fictitious magnetic field' arising
due to a `fictitious gauge potential' containing in the generalized momenta. The eigenvalues of ${\cal{M}}$
are $\alpha^2 \pm \lambda_i, i=1, 2, \dots, m$ semi-positive definite for
$\alpha^2 \geq max_i \lambda_i$. The stability properties of a system with balanced loss and gain is improved
in this region. The famous Landau problem\cite{ll} with balanced loss and gain has been studied in Ref. \cite{pkg}
with various interesting results. In particular, the representation of the matrices ${\cal{M}}, R,
{\cal{D}}$ may be considered as follows:
\be
{\cal{M}}= \frac{1}{2}
\bp
{B+C} && \gamma\\ \gamma && B-C
\ep,
R =
\bp
{0} && 1\\ -1 && 0
\ep,
{\cal{D}}= \frac{1}{2}
\bp
{-\gamma} && B+C\\ -(B-C) && \gamma .
\ep
\label{landau-rep}
\ee
\noindent for the Hamiltonian $H$ in Eq. (\ref{hami}) with $N=2$ and $V=0$. The system is described
by the equations of motion,
\bea
&& \ddot{x}_1 + \gamma \dot{x}_1 - B \dot{x}_2 - C \dot{x}_2=0\nonumber \\ 
&& \ddot{x}_2 - \gamma \dot{x}_2 + B \dot{x}_1 - C \dot{x}_1=0
\label{lan-equation}
\eea
\noindent where $B$ is the external uniform magnetic field, $\gamma$ is the loss-gain parameter and $C$
is the strength of non-Lorentzian velocity mediated coupling which appears in the study of synchronization
of coupled oscillators\cite{nli}. The eigenvalues of the matrix ${\cal{M}}$ are,
\be
\lambda_{\pm} = \frac{1}{2} \left ( B \pm \bigtriangleup \right ), \ \
\bigtriangleup \equiv \sqrt{ C^2 + \gamma^2}.
\ee
\noindent The condition for a positive-definite ${\cal{M}}$ for $C=0$ may be interpreted as 
the situation in which the magnitude of the external magnetic field $B$ supersedes the magnitude of the
`fictitious magnetic field' $\gamma$, i.e. ${\vert B \vert}> {\vert \gamma \vert}$. The matrix ${\cal{M}}$ is
singular for $B=\bigtriangleup$ and the parameter-space may be divided into three disjoint sectors:\\
(i) Region-I ( {\bf $B > \bigtriangleup$} ): The Hamiltonian is bounded from below
and the system admits periodic solutions with a reduced cyclotron frequency $\omega=\sqrt{B^2-\bigtriangleup^2}$ compared
to the standard Landau system with $\omega_L=B$.\\
(ii)  Region-II ( {\bf $ -\bigtriangleup < B < \bigtriangleup$}): The solutions are unbounded.\\
(iii)  Region-III ( {\bf $ B < - \bigtriangleup$}): The Hamiltonian is bounded from above and
in fact, it is identical with the Hamiltonian in Region-I except for an overall multiplication factor of $-1$.
The system admits periodic solutions.\\ 
The system does not admit any periodic solution for a vanishing Lorentz interaction, while periodic solutions are allowed for
$B \neq 0$. The quantum Hamiltonian admits bound states in Region-I and III. Moreover, the quantum bound states are well defined on the real line
without the need of defining the problem on any suitable Stokes wedges. The same feature persists for other Hamiltonian system
with balanced loss and gain. The details of classical and quantum solutions of the Landau Hamiltonian with balanced loss and gain along with Hall effect and
underlying supersymmetry is described in Ref. \cite{pkg}.

A few comments are in order before the end of the section.
\begin{itemize}
\item The Bateman oscillator with a modified kinetic energy term has been considered in Ref. \cite{bc} in a different context.
The system, when considered on a commutative space and cast into the notation of the present article, describes
the model of Ref. \cite{cmb} with Lorentz interaction. It can be shown that classical periodic solutions
are obtained for an extended region of parameter-space compared to that of Ref. \cite{cmb}. Further, the
quantum problem is well defined on the real line.

\item The gyroscopic force, which has the same form as Lorentz force, has been used for long to control dissipative
induce instabilities in classical systems and there are plenty of important theorems and results\cite{marsden}.
The relevant results as appropriate to a system with balanced loss and gain may be used successfully.
\end{itemize}
The inclusion of Lorentz interaction in order to stabilize a classical system with balanced loss and gain as well as well
as to define the quantum problem on the real line may be utilized for practical applications.

\section{Solvable Models}\label{sm}

Several solvable models with balanced loss and gain have been considered for
$N \geq 2$ degrees of freedom in Refs. \cite{cmb,sg-calo,bend-2,pkg-ds,sg-calo-cor,ds-pkg,pkg}.
A few solvable models are discussed below to highlight the general features of these solvable models.
The case of uniform external magnetic field and constant loss-gain terms has been discussed along
with its physical relevance in Sec. \ref{header-li}. The space-dependent loss-gain terms make
the equations highly nonlinear and in general, solvable models are rare. Nevertheless, a few solvable
models with space-dependent loss-gain terms are known\cite{ds-pkg} which will not be discussed in
this article. A few solvable models with vanishing Lorentz interaction and constant loss-gain terms
are presented below, the details of which may be found in Ref. \cite{pkg-ds}.
This corresponds to $\alpha=0$ and $F_i=x_i$ in the representation of matrices
in Sec. \ref{header-mat}. The matrices have the following expressions,
\bea
M=I_m \otimes \sigma_1, \ R=-i \gamma I_m \otimes \sigma_2, \ D_S=\gamma I_m \otimes \sigma_3
\label{mat-mat}
\eea
\noindent and $\hat{O}=\frac{1}{\sqrt{2}} \left [ I_m \otimes \left ( \sigma_1+\sigma_3 \right )
\right ]$ diagonalizes $M$ to $M_D=I_m\otimes \sigma_3$.
The orthogonal transformation generates the following set of new co-ordinates:
\be
z_i^{-}= \frac{1}{\sqrt{2}} \left ( x_{2i-1}-x_{2i} \right ),
\ \ z_i^{+}=\frac{1}{\sqrt{2}} \left ( x_{2i-1}+x_{2i} \right ), \ i=1, 2,\dots m
\label{trans}
\ee
\noindent The Eqs. of motion reads,
\be
\ddot{z}_i^{+} -2 \gamma \dot{z}_i^{-} +
 2\frac{\partial V}{\partial z_i^+}=0,\ \
\ddot{z}_i^{-} - 2 \gamma \dot{z}_i^{+} -
2 \frac{\partial V}{\partial z_i^-}=0,\ \ i=1, 2, \dots m.
\label{z-eqn}
\ee
\noindent In the subsequent discussions $2V$ is replaced by $V$.
The Eqs. (\ref{z-eqn}) is a set of $2m$ coupled differential equations which
take simple form for (i) translational and (ii) rotational invariance of the
system and solvable for specific choices of potential.

\subsection{Translational invariance}

The choice of the potential $V\equiv V(z_i^-)$ or $V\equiv V(z_i^+)$ allows
a decoupling of Eq. (\ref{z-eqn}). The potential $V(z_i^-)$ remains invariant
under the translations $x_{2i-1} \rightarrow x_{2i-1} +
\eta_i, x_{2i}\rightarrow x_{2i} + \eta_i$, where $\eta_i$'s are $m$
independent parameters. The form of the potential is special in the sense that
it allows $m$ independent parameters $\eta_i$ instead of a single one.  The
translational invariance leads to $m$ integrals of motion $\Pi_i$ which are
in involution:
\be
\Pi_i= 2P_{z_i^+} -\gamma z_i^-, \ \{\Pi_i,\Pi_j\}_{PB}=0, \ \ \{H, \Pi_i\}_{PB}=0,
\ee
\noindent where $\{,\}_{PB}$ denotes the Poisson bracket and $P_{z_i^+}$
is the conjugate momenta corresponding to $z_i^+$. The Hamiltonian along with
$\Pi_i$'s constitute $m+1$ integrals of motion for a system with $N=2m$ particles,
implying that the system is at least partially integrable for $N > 2$ and completely
integrable for $N=2$.  A similar analysis can be performed for $V(z_i^+)$. The
discussion in this article is restricted to the case $V\equiv V(z_i^-)$. 

With the introduction of new co-ordinates $z_i= z_i^- + \frac{\Pi_i}{2 \gamma}$,
Eq. (\ref{z-eqn}) can be rewritten as,
\bea
&& \ddot{z}_i - 4 \gamma^2 {z}_i -  \frac{\partial V(z_i)}{\partial z_i}= 0,\
\nonumber \\
&& z_i^+(t) = 2 \gamma \int z_i(t) dt + C_i, \ \ i=1, 2, \dots m,
\label{homo-eqn}
\eea
\noindent where $C_i$ are $m$ integration constants. The potential depends
on the gain-loss parameter via $z_i$ which may be avoided by choosing the
constants of motion $\Pi_i=0 \ \forall \ i$ leading to the initial conditions
$\dot{z}_i^+(0) = 2 \gamma z_i^-(0) \ \forall \ i$, where $\dot{z}_i^+(0)$
and $z_i^-(0)$ can be chosen depending on the physical requirements.
The problem now lies  to find $V$ for which the first equation of
Eq. (\ref{homo-eqn}) is exactly solvable. Several examples involving coupled
chain of cubic oscillators, potential solely dependent on the radial variable
$r=\sum_{i=1}^m z_i^2$ in the sub-system defined by the co-ordinates 
$(z_1, z_2, \dots, z_m)$, Calogero-type inverse square interaction, and
Henon-Heils potential are considered in Ref. \cite{pkg-ds}. The case of
cubic oscillator is discussed below.

The choice of $V$ for the simplest case of $N=2,m=1$ which produces a
solvable system is the following:
\bea
&& V(z_1) =-2 \omega_0^2 z_1^2 - \frac{\alpha}{4} z_1^4, \ \omega, \alpha \in
\Re,\nonumber \\
&& \ddot{z}_1 + \omega^2 {z}_1 + \alpha z_1^3=0, \
\omega^2 \equiv 4 ( \omega_0^2-\gamma^2 ).
\label{cubic-o}
\eea
\noindent There are three distinct regions in the parameter-space for which non-singular stable solutions can be
obtained analytically in terms of Jacobi Elliptic functions \textemdash
Region-I: $\omega^2 > 0, \alpha > 0$,  Region-II: $\omega^2 > 0, \alpha < 0$ and
Region-III: $\omega^2 < 0, \alpha > 0$. The stability requires $-\omega_0 < \gamma < \omega_0$ for
region-I and Region-II, while $-\gamma < \omega_0 < \gamma$ in region-III. The nonlinear interaction allows
$\gamma > \omega_0$ which has not been seen for system with linear interaction. There is an additional
constraint in each region involving the amplitude and frequency of the solution, and the nonlinear coupling $\alpha$
for the existence of non-singular stable solution. For example, the solution in Region-I,
\bea
&& z_1(t)= A \ cn (\Omega t,k),\ \
z_1^+(t) = \frac{2 \gamma }{\Omega} \frac{\cos^{-1} \{ dn(\Omega t,k) \}
sn(\Omega t,k)}{\sqrt{1-dn^2(\Omega t,k)}}\nonumber \\
&& \Omega=\sqrt{\omega^2 + \alpha A^2}, \
k^2=\frac{\alpha A^2}{2 \Omega^2},
\label{solex1}
\eea
\noindent is non-singular and stable for $-\omega_0 < \gamma < \omega_0$ and $0 < k < 1$. The solutions
in other regions are given in Ref. \cite{pkg-ds}. The solvable models for higher $N$ can be constructed
by using the known results of coupled cubic oscillators\cite{sahadev}. The case of $N=4, m=2$ along with
its exact, non-singulars and stable solutions are also discussed in Ref. \cite{pkg-ds}. 

\subsection{Rotational Invariance}

The Bateman oscillator admits a constant of motion in addition to the Hamiltonian $H_B$,
\bea
L_B= x \dot{y} - y \dot{x} + 2 \gamma xy
\eea
\noindent thereby making it an integrable system. Their exist  $m$ constants of motion similar to $L_B$ for
a system with $N=2m$ particles with the matrices ${\cal{M}}, R, {{D}}_S$ given in Eq. (\ref{mat-mat}) and
potential of specified form.
In particular, $\hat{L}=\frac{1}{2} \left ( P^T D_S X + X^T D_S P \right ) $ is a constant of motion for
potential satisfying the condition $X^T D_S \frac{\partial V}{\partial X}=0$  which may be solved with the
ansatz,
\bea
V \equiv  V(\tilde{r}), \ \tilde{r}^2 \equiv X^T G X, \{G,D_s\}=0
\eea
\noindent where $G$ is a symmetric matrix and $D_S$ need not be restricted to any specific representation.
For $N=2$ and with the representation of matrices given by Eq. (\ref{mat-mat}), $G$ is uniquely determined
as $G=M=\sigma_1$. However, for $N > 2$, the choice of $G$ is not unique and there are several possibilities\cite{pkg-ds}.
The variable $\tilde{r}$ takes the form 
\be
\tilde{r}^2 = \sum_{i=1}^{m} x_{2 i-1} x_{2 i} = \sum_{i=1}^m \left [ \left ( z_i^+
\right )^2 - \left ( z_i^- \right )^2 \right ],
\label{N-radi}
\ee
\noindent for $G=M^{-1}$ and representation of matrices given by Eq. (\ref{mat-mat}).
It may be noted that $\tilde{r}$ has the interpretation of the radial
variable in a pseudo-Euclidean co-ordinate with the signature of the metric
as $(1,-1,1,-1, \dots, 1, -1)$. The $m$ constants of motion in addition to the Hamiltonian have
the expressions,
\be
\hat{L} _i = \dot{z}_i^+ z_i^- - z_i^+ \dot{z}_i^- + \gamma \left [ ( z_i^+)^2
-(z_i^-)^2 \right], \ i=1, 2, \dots, m,
\ee
\noindent which are in involution,
\be
\{\hat{L} _i, \hat{L} _j \}_{PB} =0, \ \{H, \hat{L} _i\}_{PB}=0,
\ee
\noindent implying that the system is at least partially integrable for $N > 2$
and integrable for $N=2$ for a system of $N=2m$ particles.
The values of all the constants of motion are zero, i.e.
$\hat{L}_i=0 \ \forall \ i$ and $r^2=\sum_{i=1}^m q_i^2\equiv q^2$ for the
following parameterizations of the co-ordinates,
\be
z_i^+(t) = q_i(t) \cosh(\gamma t), \
z_i^-(t) = q_i(t) \sinh(\gamma t)
\ee
\noindent The equations of motion can be expressed solely in terms of $q_i$,
\be
\ddot{q}_i - \gamma^2 q_i +
\frac{1}{q} \frac{\partial V(q)}{\partial q} q_i=0,
i=1, 2, \dots, m,
\label{qq-eqn}
\ee
\noindent and exactly solvable models may be constructed for suitable choices of $V(q)$.
For example, $V(q)=\frac{1}{2} \omega^2 q^2 + \frac{\alpha}{4} q^4$ leads to an exactly
solvable equation\cite{sahadev},
\be
\ddot{q}_i +  \Omega^2 q_i + \alpha q^2 q_i=0, \ \Omega^2=\omega^2 - \gamma^2.
\label{hyperbolic}
\ee
\noindent The Bateman oscillator is reproduced for $\alpha=0, N=2$ for which the two modes are not coupled.
The nonlinear interaction couples all the modes and the corresponding Hamiltonian is not separable
like Bateman oscillator. Although the above equation (\ref{hyperbolic}) admits non-singular stable solutions, when
expressed in terms of the original variables $x_i$, there are growing as well as decaying modes
\textemdash $x_{2i}$ are always decaying in time, while that of $x_{2i-1}$ grows with time.
The situation can not be saved by allowing $\Omega$ to be complex, i.e ${\vert \gamma \vert} >
{\vert \omega \vert}$, since the decay and growth of the solutions are of the form $e^{\pm \gamma t}$
which is independent of $\Omega$.
The volume of the flow in the phase-space is always preserved. The same behaviour has been seen for
the Bateman oscillator, where the two modes are decoupled and there is no nonlinear interaction. However, inclusion
of coupling among different co-ordinates via nonlinear interaction does not change the situation
for $\hat{L}_i=0$. The solutions for $\hat{L}_i \neq 0$ may give stable solutions.
The addition of a term that breaks the rotational invariance may also allow stable solutions. 
Different quantization schemes as applied to Bateman oscillator have revealed a host of intricate
issues\cite{morse,bopp,feshback,trikochinsky,dekker-0,rasetti,rabin,jur,bc}, the
quantization of the quartic oscillator corresponding to the $V(q)$ deserves further
attention.

\section{Role of ${\cal{PT}}$-symmetry}\label{hpt}

The discrete symmetries like parity(${\cal{P}}$), time-reversal(${\cal{T}}$) and
charge-conjugation(${\cal{C}}$) play an important role in physics. The ${\cal{PT}}$-symmetric
quantum mechanics is one such area where non-relativistic non-hermitian Hamiltonian
with ${\cal{PT}}$ symmetry admit entirely real spectra in the unbroken ${\cal{PT}}$-regime\cite{bbook}.
In general, the eigen-states are not orthogonal and the unitary time-evolution is not possible. However,
with a modified ${\cal{CPT}}$ norm, a consistent quantum description with unitary time-evolution is allowed\cite{bbook}.
There is no concept of anti-particles in non-relativistic quantum mechanics. The charge conjugation operator
${\cal{C}}$ is introduced with the expectation that at a more fundamental level it will be related to
the ${\cal{CPT}}$ invariance of any Lorentz invariant local quantum field theory. It may be noted that the
${\cal{CPT}}$ theorem\cite{green}, which was originally derived for a hermitian Hamiltonian, has now
been extended to the case of non-hermitian Hamiltonian\cite{pdm}. There is no signature of
violation of Lorentz invariance in nature so far and the ${\cal{CPT}}$ is considered to be a fundamental
symmetry of nature.

The studies on ${\cal{PT}}$-symmetric systems have been diversified to many areas, including the obvious choice
of ${\cal{PT}}$-symmetric classical systems. If the operator ${\cal{C}}$ of ${\cal{CPT}}$-norm of the
${\cal{PT}}$-symmetric quantum mechanics is expected to be related to the charge-conjugation operator of Lorentz invariant
local quantum field theory, then the parity and the time-reversal operation for the {\it{classical non-relativistic}}
${\cal{PT}}$-{\it{symmetric system}} should necessarily be described by a linear transformation, since the Lorentz
transformation itself is linear. Within this background, the time-reversal transformation ${\cal{T}}$ and the most general
parity transformation ${\cal{P}}$ in two space dimensions may be defined as,
\bea
&&  {\cal{T}}: t \rightarrow -t, \ \tilde{P}_x \rightarrow - \tilde{P}_x, \
\tilde{P}_y \rightarrow - \tilde{P}_y \nonumber \\
&& {\cal{P}}:
\begin{pmatrix}
{x}\\ {y}
\end{pmatrix} \rightarrow
\begin{pmatrix}
{x^{\prime}}\\ y^{\prime}
\end{pmatrix}
=
\begin{pmatrix}
{x \cos \theta + y \sin \theta}\\
{x \sin \theta - y \cos \theta}
\end{pmatrix}, \
\begin{pmatrix}
{\tilde{P}_x}\\ {\tilde{P}_y}
\end{pmatrix} \rightarrow
\begin{pmatrix}
{\tilde{P}_x^{\prime}}\\ \tilde{P}_y^{\prime}
\end{pmatrix}
=
\begin{pmatrix}
{\tilde{P}_x \cos \theta + \tilde{P}_y \sin \theta}\\
{\tilde{P}_x \sin \theta - \tilde{P}_y \cos \theta}
\end{pmatrix}
\label{pt-tran}
\eea
\noindent where $\theta \in (0,2\pi)$. The Bateman oscillator in Eq. (\ref{bo}) is invariant under ${\cal{PT}}$ symmetry
for two values of $\theta$, namely $\theta=\frac{\pi}{2}, \frac{3 \pi}{2}$.  However, the classical solutions are not
invariant under ${\cal{PT}}$ symmetry, i.e. ${\cal{PT}} X \neq \pm X$ in any region of the parameter-space.
Thus, there is no ${\cal{PT}}$-symmetric phase for the Bateman oscillator. Consider 
Eq. (\ref{bo2}) with $G_1(x,y)=\epsilon y, G_2(x,y)=\epsilon x$ which describes the model considered in \cite{cmb}.
The system is ${\cal{PT}}$-symmetric and admits periodic solutions in certain regions in the parameter-space. There are
both ${\cal{PT}}$-broken and ${\cal{PT}}$-unbroken regions for this model. The existence of periodic solutions
in the ${\cal{PT}}$-unbroken phase lead further investigations on ${\cal{PT}}$-symmetric system with balanced loss and gain
in various directions\cite{sg-calo,bend-2,igor,khare,pkg-ds,sg-calo-cor,ds-pkg,pkg} by including nonlinear
interaction\cite{sg-calo,igor,khare,pkg-ds,ds-pkg}, many-particle system\cite{bend-2,pkg-ds,sg-calo-cor}, Lorentz
interaction\cite{pkg}, etc. The general result common to all these systems is that periodic solutions are obtained
in the unbroken ${\cal{PT}}$ phase of a ${\cal{PT}}$-symmetric system.

It has been shown recently that non-${\cal{PT}}$ symmetric classical system may also admit periodic
solutions in a coupled Duffing-oscillator with balanced loss and gain\cite{pkg-pr}. This is a new result with far reaching
consequences. Many non-${\cal{PT}}$ symmetric system with balanced loss and gain may be included in the mainstream
investigations which may exhibit novel features including existence of periodic solutions. It should be
mentioned here that there exist many other non-${\cal{PT}}$-symmetric systems with balanced loss and gain which admit periodic
solutions in certain regions of the parameter space\cite{pr-pkg}. The examples include mechanical systems with
finite degrees of freedom, models of dimer, non-relativistic field theory etc., which can be broadly classified
into Hamiltonian and non-Hamiltonian systems. A few examples are described below.

\subsection{Hamiltonian System}

\subsubsection{A coupled  Duffing Oscillator model:}

The equations of motion of the system are,
\bea
&& \ddot{x} + 2\gamma\dot{x}+\omega^{2}x+\beta_{1}y+ g x^{3}=0,\nonumber \\
&& \ddot{y} - 2\gamma\dot{y}+\omega^{2}y+\beta_{2}x+3 g x^{2}y=0
\label{v0-duff-eqn}
\eea
\noindent The linear coupling between the two modes are asymmetric for $\beta_1 \neq \beta_2$.
The system reduces to the one considered in Ref. \cite{cmb} for $g=0, \beta_1=\beta_2$. The
equation of motion for $x$ degree of freedom decouples for $\beta_1 =0$ and describes a damped
Duffing oscillator, while the $y$ degree of freedom is unidirectionally coupled to $x$. There is
no explicit forcing term. However,  the term $\beta_1 y$ in the first equation of
(\ref{v0-duff-eqn}) acts as a driving term for the damped Duffing oscillator in a non-trivial
way. The system also admits a Hamiltonian,
\bea
H_D = P_x P_y + \gamma \left ( y P_y - x P_x \right ) + \left ( \omega^2-
{\gamma^2} \right ) x y + \frac{1}{2} \left ( \beta_2 x^2 +
\beta_1 y^2 \right ) + g x^3 y,\
\label{hami-1}
\eea
\noindent where $P_x=\dot{y} - \gamma y$ and $P_y=\dot{x} + \gamma x$
are canonical momenta. In general, the system is non-${\cal{PT}}$-symmetric for the ${\cal{P}}$ and ${\cal{T}}$
defined in Eq. (\ref{pt-tran}). The first, second and the third terms in $H_D$ are invariant under
${\cal{PT}}$ only for $\theta=\frac{\pi}{2}, \frac{3 \pi}{2}$. The fourth and the fifth terms
representing linear asymmetric coupling and nonlinear interaction, respectively, are not invariant
under ${\cal{PT}}$ symmetry. It can been shown that for vanishing nonlinear coupling,
i.e. $g=0$, the corresponding non-${\cal{PT}}$-symmetric system with asymmetric linear coupling
does not admit any periodic solutions. This is consistent with the general folklore that
a system with balanced loss and gain admits periodic solution only for a ${\cal{PT}}$-symmetric
system. However, although the system is non-${\cal{PT}}$ symmetric for $g \neq 0, \beta_1=\beta_2$, it
admits periodic solutions\cite{pkg-pr} in some regions in the parameter-space. 

The independent scales in the system may be fixed by employing the following transformations,
\bea
t \rightarrow \omega^{-1} t, \ x \rightarrow {\vert \beta_2
\vert}^{-\frac{1}{2}} x, \
y \rightarrow {\vert \beta_1 \vert}^{-\frac{1}{2}} y, \ \beta_1 \neq 0,  \beta_2 \neq 0,
\label{scale}
\eea
\noindent  which allows a reduction in total number of independent parameters which is
convenient for analyzing the system. The model can be described in terms of
three independent parameters $\Gamma, \beta$ and $\alpha$ defined as,
\bea
\Gamma = \frac{\gamma}{\omega}, \beta= \frac{\sqrt{{\vert {\beta_1} \vert}
{\vert {\beta_2}\vert}}}{\omega^2},
\alpha= \frac{g}{{\vert \beta_2 \vert} \omega^2},
\eea
\noindent and the equations of motion have the following expressions:
\bea
&& \ddot{x} + 2 \Gamma \dot{x}+ x + \sgn(\beta_1) \ \beta y + \alpha x^{3}=0, \nonumber \\
&& \ddot{y} - 2 \Gamma \dot{y}+ y + \sgn(\beta_2) \ \beta x+ 3 \alpha x^{2}y=0.
\label{duff-eqn}
\eea
The limit to the linear system $g \rightarrow 0$ now corresponds to $\alpha \rightarrow 0$.
The system admits regular periodic as well as chaotic solutions for the case
$\sgn(\beta_1) \sgn(\beta_2)=1$ and the discussion is restricted to $\sgn(\beta_i)=1, i=1,2$.
The results for $\sgn(\beta_i)=-1$ can be obtained from that of results for $\sgn(\beta_i)=1$
by simply allowing $\beta \rightarrow - \beta$.

\begin{figure}[h!]
\begin{minipage}{3in}
\centering
\includegraphics[width=3in]{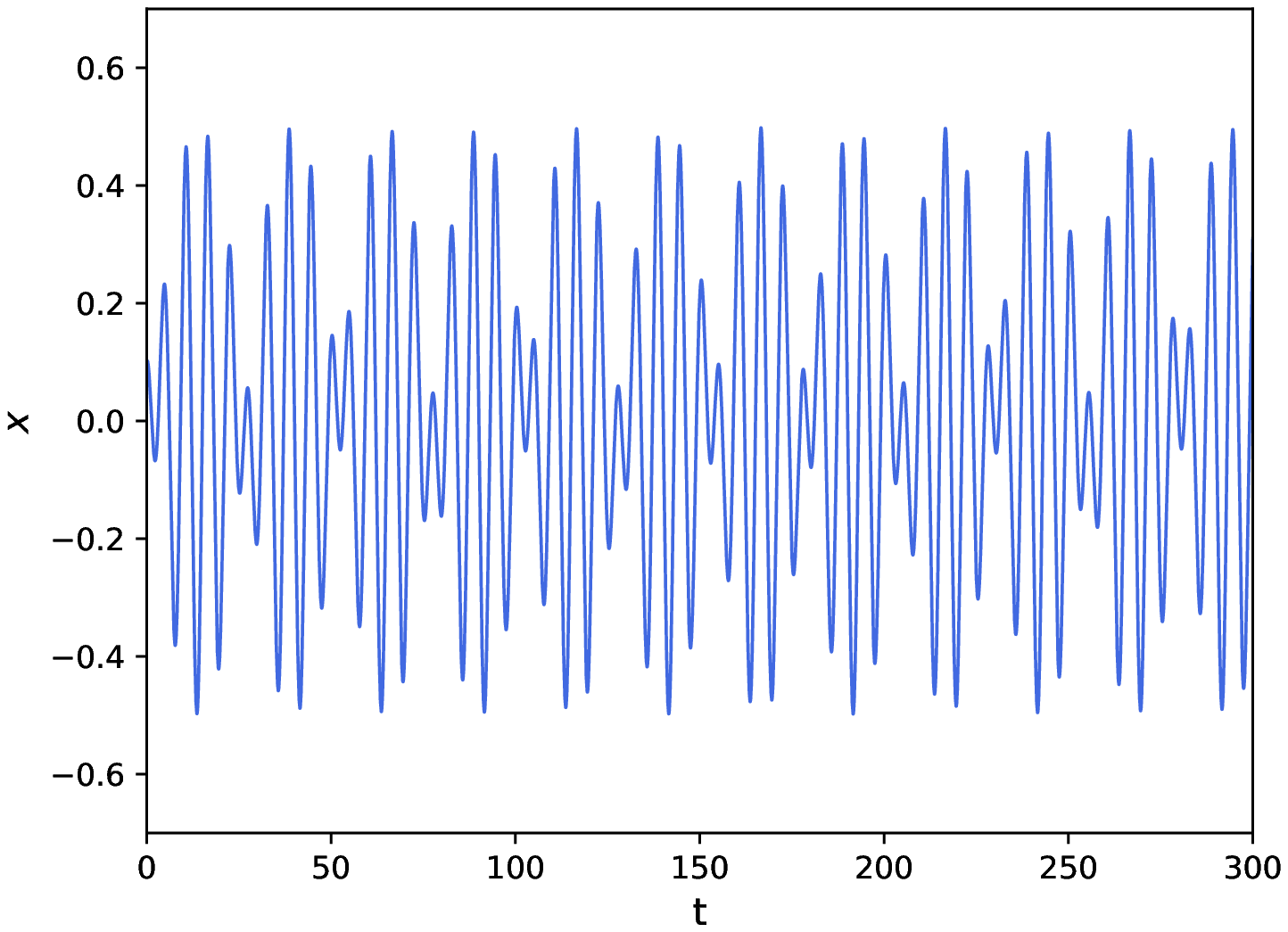}
\label{xtimeseries1}
\end{minipage} \hspace{.2in}%
\begin{minipage}{3in}
\centering
\includegraphics[width=3in]{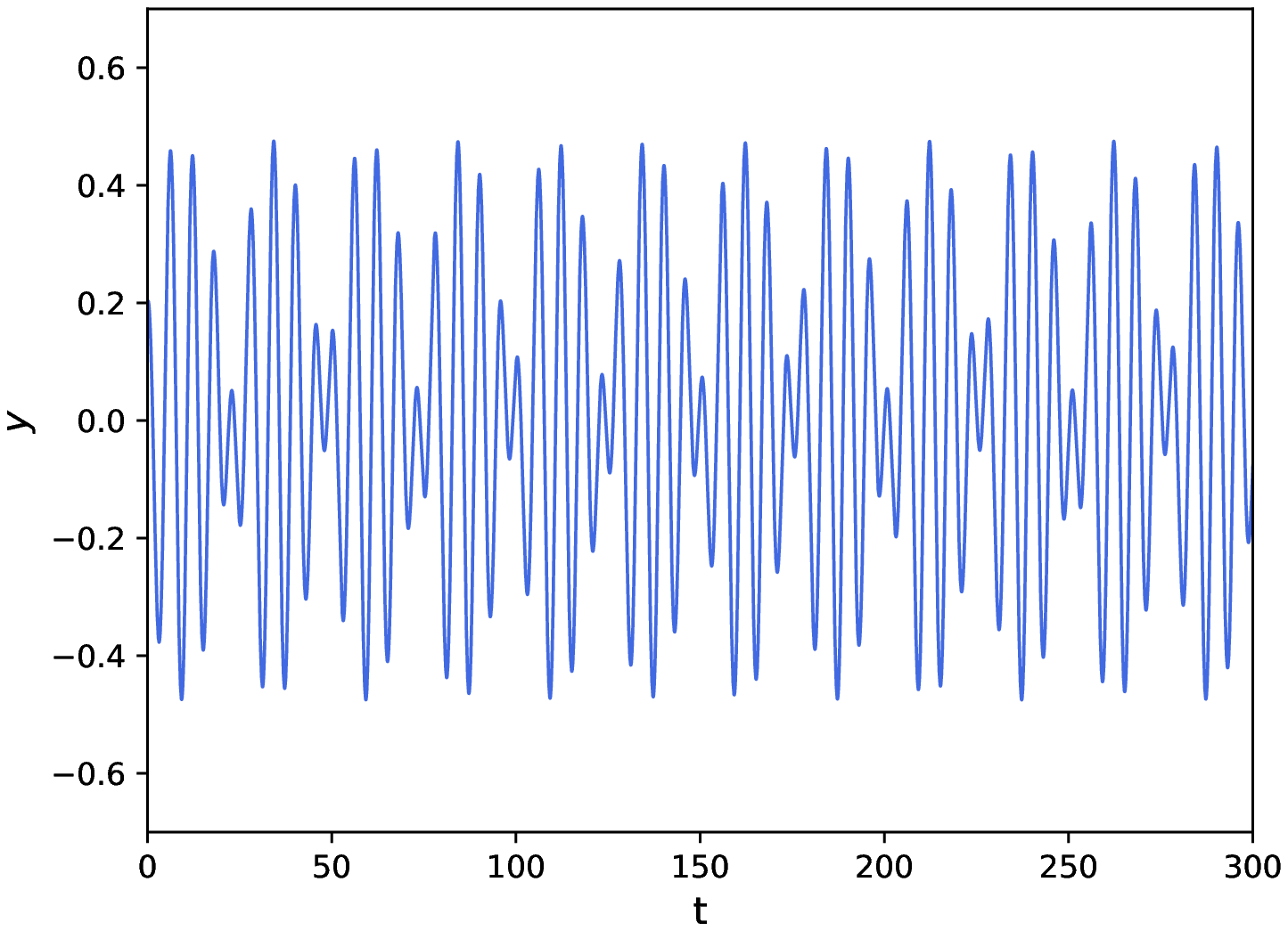}
\end{minipage}
\caption{(Color online)
Regular solutions of Eq. (\ref{duff-eqn}) in the vicinity of the point
$P_0$ with the initial conditions $x(0)=.1, y(0)=0.2, \dot{x}(0)=.03$, $\dot{y}(0)=.04$ and
$\alpha=1, \beta=.5, \Gamma=.2$. ({\it Reproduced from Ref. \cite{pkg-pr}})}
\label{p0ts}
\end{figure}

\begin{figure}
\begin{minipage}{3in}
\centering
\includegraphics[width=3in]{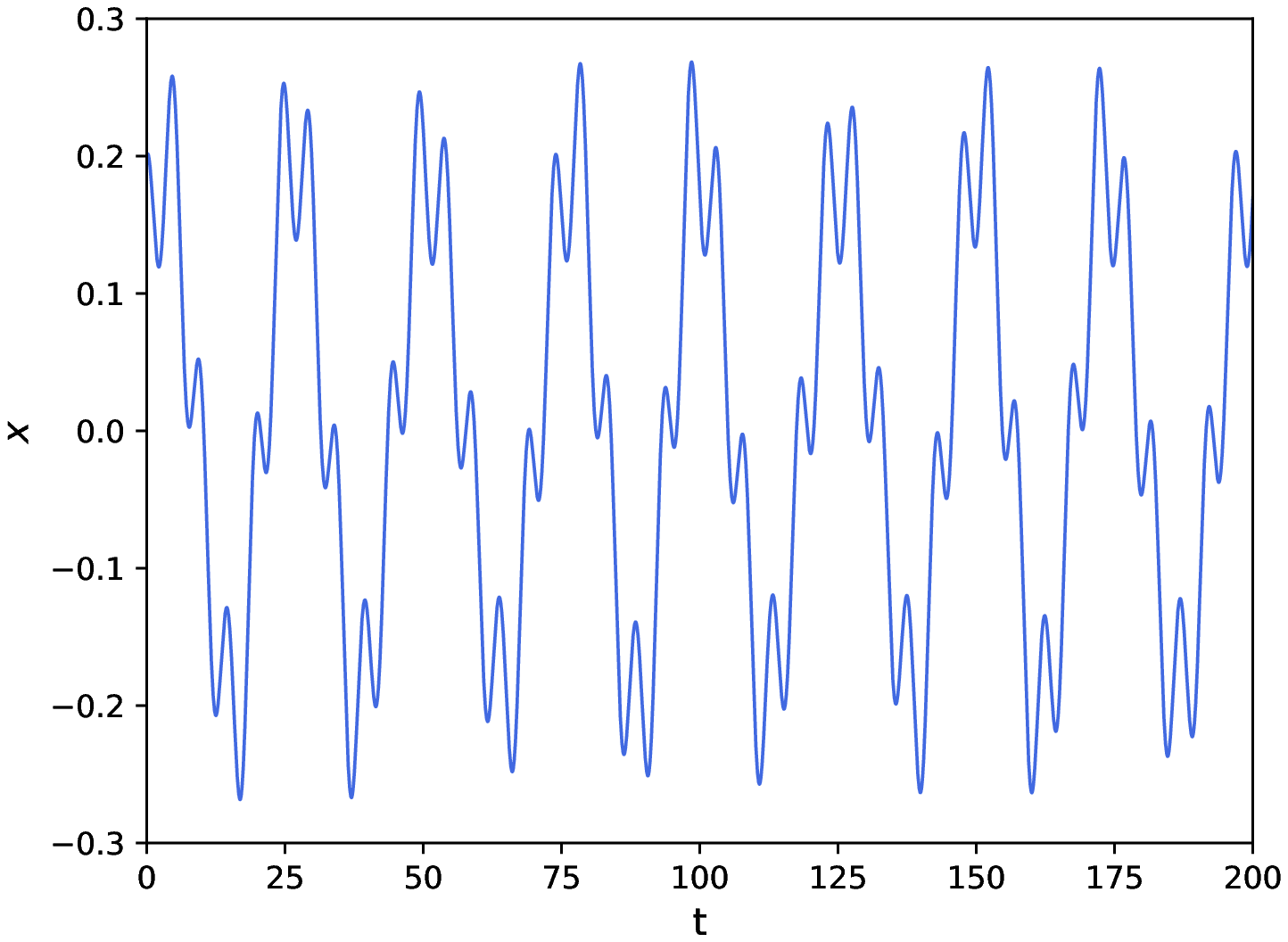}
\label{xtimeseries1}
\end{minipage} \hspace{.2in}%
\begin{minipage}{3in}
\centering
\includegraphics[width=3in]{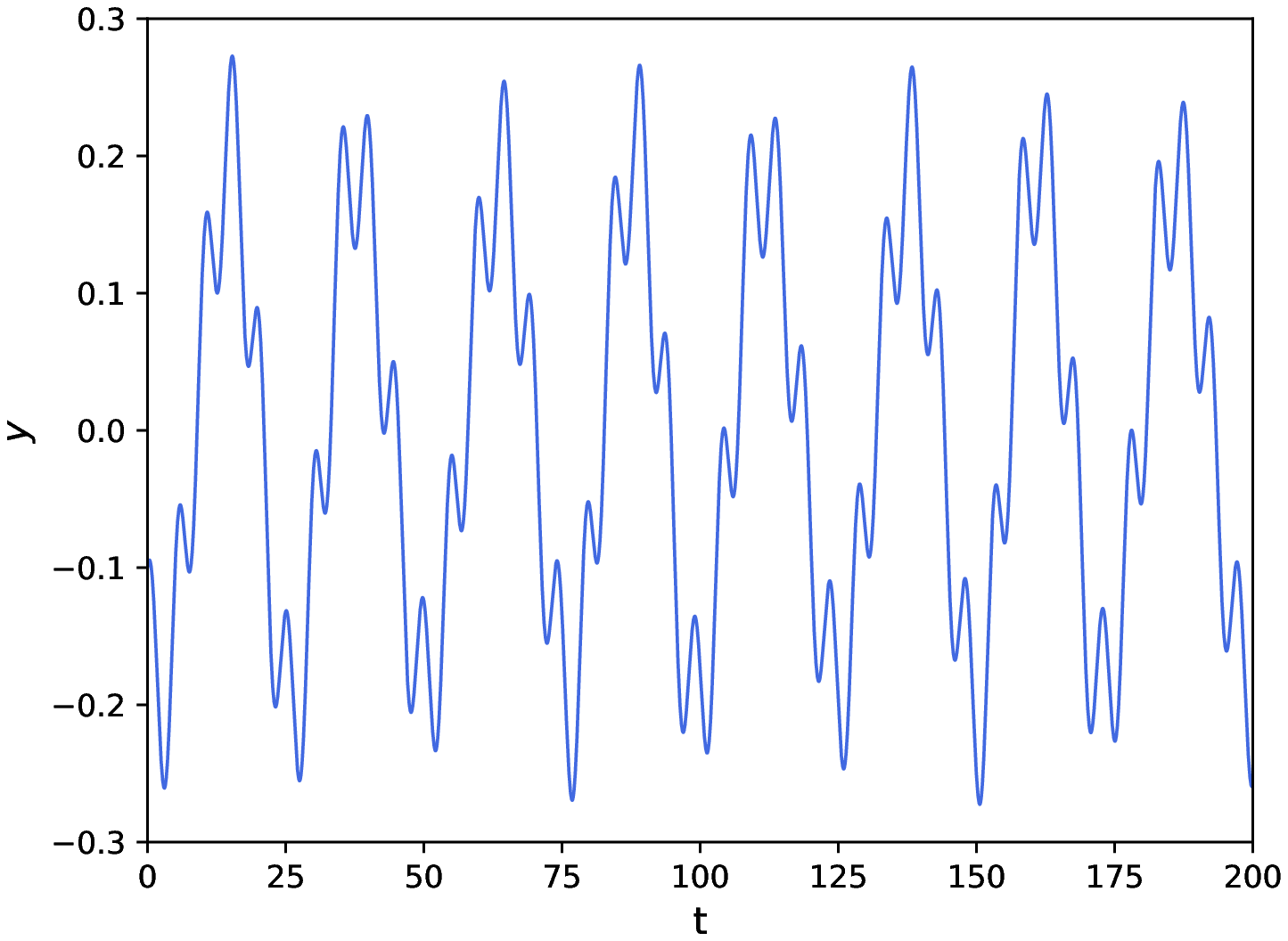}
\end{minipage}
\caption{(Color online)  Regular solutions of Eq. (\ref{duff-eqn}) in the
vicinity of the point $P_1^{+}$ with the initial conditions $x(0)=.2, y(0)=-0.1, \dot{x}(0)=.02$,
$\dot{y}(0)=.03$ and $\alpha=1, \beta=1.01, \Gamma=.3$.({\it Reproduced from Ref. \cite{pkg-pr}}) }
\label{p1ts}
\end{figure}

The system admits five equilibrium points $P_0, P_1^{\pm},
P_2^{\pm}$ in the phase-space $(x,y,\tilde{P}_{x},\tilde{P}_{y})$ of the system,
where $\tilde{P}_x = \dot{y} - \Gamma y, \ \tilde{P}_y = \dot{x} + \Gamma x$.
The equilibrium points are,
\be
P_0=(0,0,0,0), P_1^{\pm}=(\pm \delta_+,\pm\eta_+,\mp \Gamma \eta_+,\pm \Gamma
\delta_+),
P_2^{\pm}=(\pm \delta_-,\pm \eta_-,\mp \Gamma \eta_-,\pm \Gamma \delta_-),
\ee
\noindent where $\delta_{\pm}$ and $\eta_{\pm}$ are defined as follows:
\bea
\delta_{\pm}=\frac{1}{\sqrt{3 \alpha}} \left [ -2  \pm \sqrt{ 1 + 3 \beta^2}
\right]^{\frac{1}{2}}, \ \
\eta_{\pm}=-\frac{\delta_{\pm}}{3 \beta} \left [ 1 \pm \sqrt{1 + 3 \beta^2}
\right ].
\eea
\noindent The application of Dirichlet theorem\cite{dirichlet} is inconclusive. The linear
stability analysis shows that the points $P_0$ and $P_1^{\pm}$ are
stable in the following regions of the parameter space:
\bea
&& P_0: -\frac{1}{\sqrt{2}} < \Gamma < \frac{1}{\sqrt{2}}, \ \
4 \Gamma^2 \left ( 1 - \Gamma^2 \right ) < \beta^2 < 1\nonumber \\
&& P_1^{\pm}: \beta^2 >1, \ \ \Gamma^2 \leq \frac{\sqrt{2}-1}{2}.
\label{stab-con}
\eea
\noindent The points $P_2^{\pm}$ are not stable anywhere in the parameter-space. These
results are confirmed by perturbative and numerical analysis\cite{pkg-pr} and the system
admits periodic solutions around the equilibrium points. Regular solutions of Eq. (\ref{duff-eqn}) in
the vicinity of the point $P_0$ and $P_1$ are plotted in Figs. \ref{p0ts} and \ref{p1ts}, respectively
for $\alpha=1, \beta=.5, \Gamma=.2$.

\begin{figure}[h!]
\begin{minipage}{3in}
\centering
\includegraphics[width=3in]{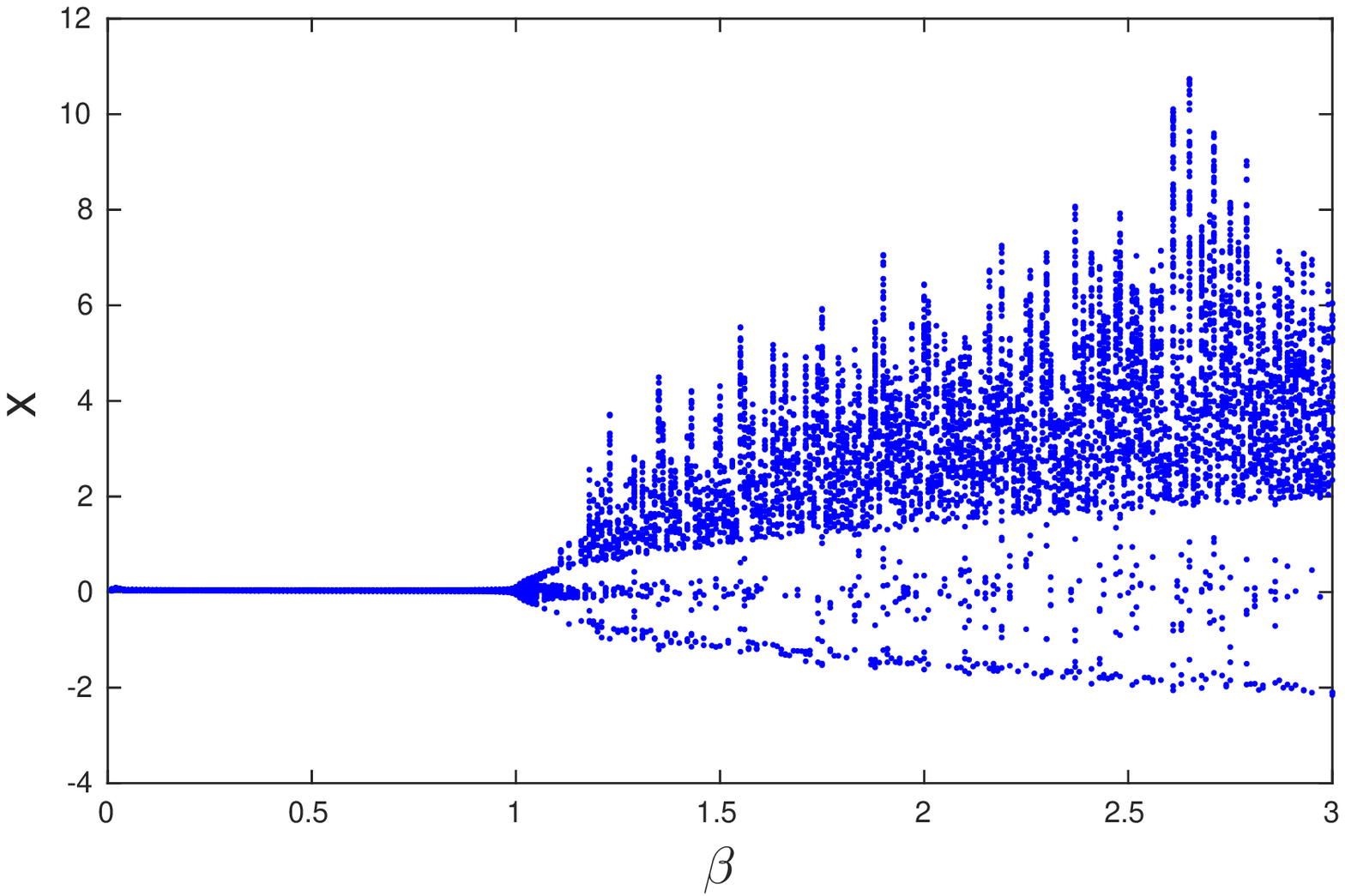}
\label{xtimeseries1}
\end{minipage} \hspace{.2in}%
\begin{minipage}{3in}
\centering
\includegraphics[width=3in]{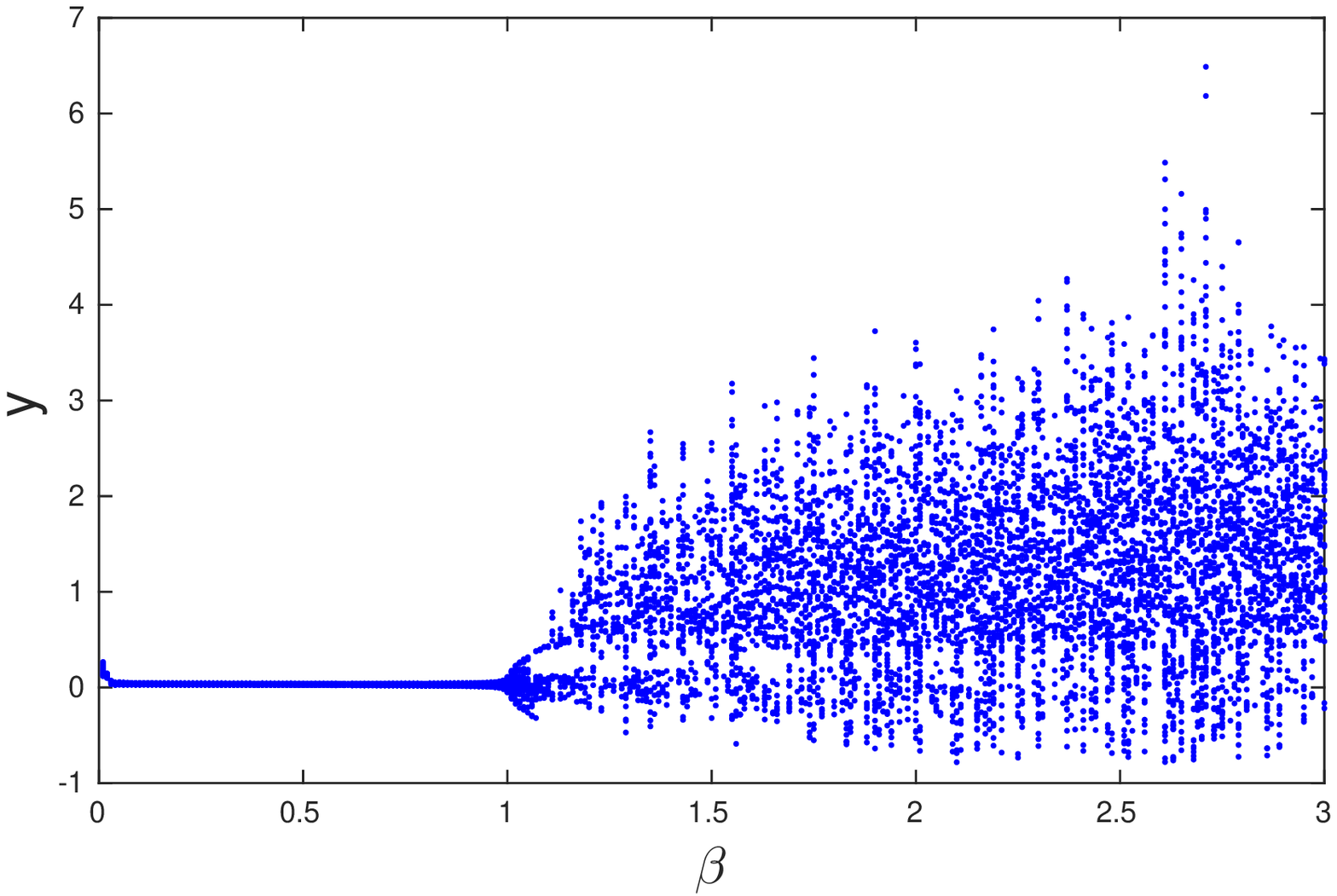}
\label{ytimeseries1}
\end{minipage}
\caption{(Color online) Bifurcation diagrams for $\beta$ with $\Gamma=0.01$ and $\alpha=.5 $
with the initial conditions $x(0)=0.01$, $y(0)=.02$, $\dot{x}(0)=.03$, $\dot{y}(0)=.04$.({\it Reproduced from Ref. \cite{pkg-pr}}) }
\label{fig-bifr}
\end{figure}

The bifurcation diagram is plotted in Fig. \ref{fig-bifr} for varying $\beta$ and
$\Gamma=0.01, \alpha=.5$. It can be
seen that the chaotic regime begins for $\beta > \beta_c \equiv 1.05$. In general,
the chaotic regime starts beyond ${\vert \beta \vert} > 1$ for a range of values of $\alpha$
and $\Gamma$. The existence of chaotic dynamics is confirmed numerically through various 
means \textemdash sensitivity of time-series to the initial conditions, Poincar$\acute{e}$ section,
power-spectra, auto-correlation functions and computation of Lyapunov exponents. The details of the
numerical investigations are given in Ref. \cite{pkg-pr}, only the plots of the  Lyapunov exponents 
for the initial condition $x(0)=.01, y(0)=.02, \dot{x}(0)=.03, \dot{y}(0)=.04$ are
shown in the left panel of the Fig. \ref{multi} which have the values $(.13248, .0015691, -.0016145, -.13244)$.
The standard result
that the sum of the Lyapunov exponents are zero for a Hamiltonian system may be verified
within the numerical approximations by taking the values of the Lyapunov
exponents up to the third decimal places with an error of the order of $10^{-4}$. 

\subsubsection{Non-${\cal{PT}}$-symmetric Positional Non-conservative force}

\begin{figure}[h!]
\begin{minipage}{3in}
\centering
\includegraphics[width=3in]{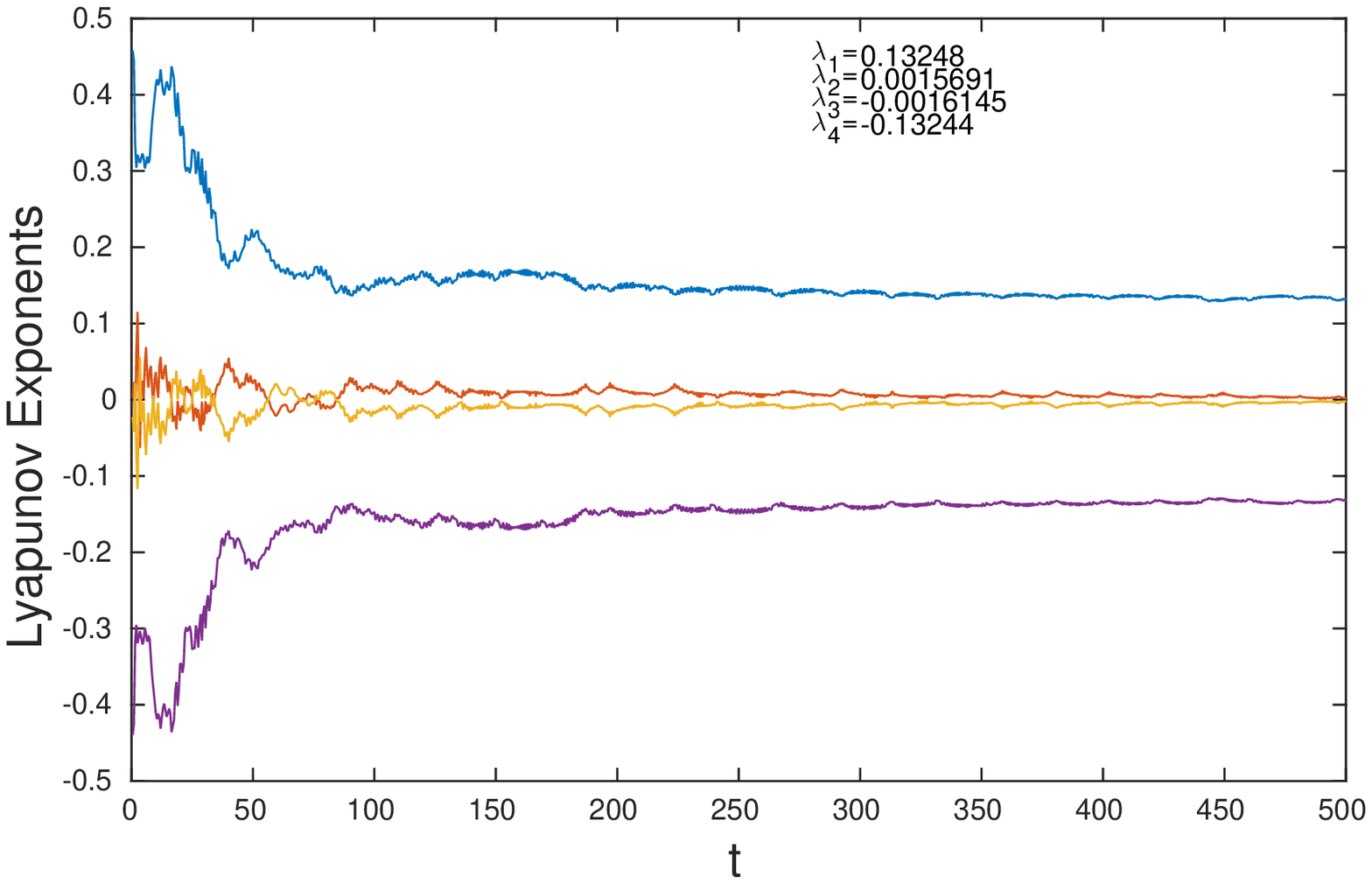}
\label{xtimeseries1}
\end{minipage} \hspace{.2in}%
\begin{minipage}{3in}
\centering
\includegraphics[width=3in]{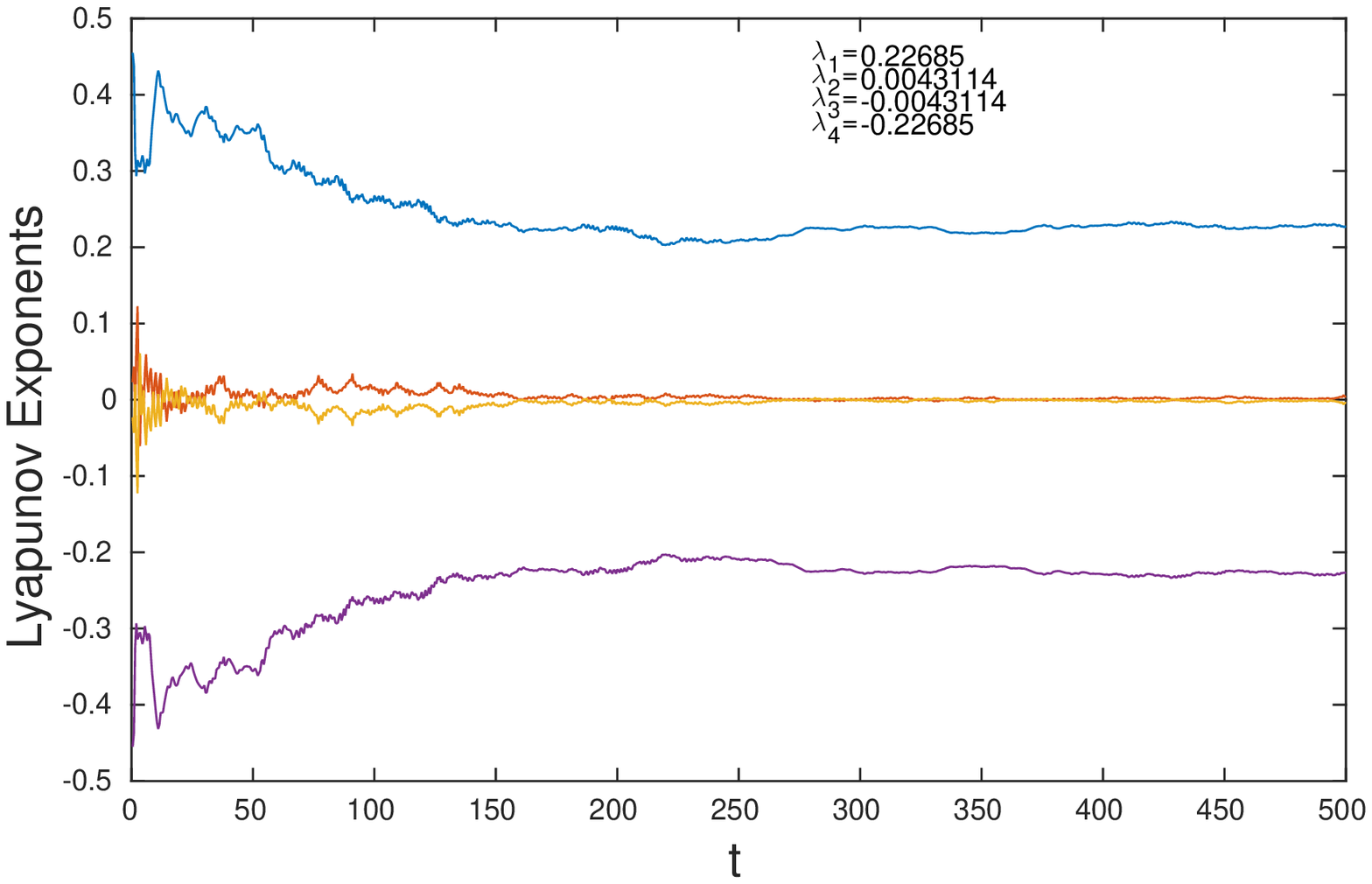}
\label{ytimeseries1}
\end{minipage}
\caption{(Color online)  Lyapunov exponents for $\beta=1.5, \alpha=.5$ and (a) $\Gamma=.01$ (left panel), (b)
$\Gamma=0$ (right panel). The initial conditions for both the cases are same,
 $x(0)=.01, y(0)=0.02, \dot{x}(0)=.03, \dot{y}(0)=.04$. ({\it Reproduced from Ref. \cite{pkg-pr}}) }
\label{multi}
\end{figure}

The Hamiltonian $H_D$ shows chaotic behaviour even for $\Gamma=0$, i.e. the system without the
vanishing loss-gain terms, thereby providing an example of Hamiltonian chaos for two undamped Duffing
oscillators coupled to each other in a specific way. In particular, the Hamiltonian $H_D$ can be
rewritten for $\Gamma=0$ as,
\bea
H_D =  \left ( \frac{1}{2} P_u^2+ \frac{\Omega_+}{2} u^2 +\frac{\alpha}{4} u^4 \right ) -
\left ( \frac{1}{2} P_v^2 + \frac{\Omega_-}{2} v^2 + \frac{\alpha}{4} v^4 \right )
+ \frac{\alpha \ u v}{2} \left ( u^2-v^2 \right ),  \Omega_{\pm} =1 \pm \beta,
\eea
\noindent where the new co-ordinates and momenta are defined as,
\bea
u=\frac{x+y}{\sqrt{2}}, \ v=\frac{x-y}{\sqrt{2}}, \ P_u=\frac{\tilde{P_x}+\tilde{P}_y}{\sqrt{2}}, \
P_v=\frac{\tilde{P_x}-\tilde{P}_y}{\sqrt{2}}.
\eea 
\noindent This particular representation of $H_D$ describes two undamped Duffing oscillators,
with different angular frequencies corresponding to the harmonic terms and identical nonlinear terms,
coupled to each other through specified nonlinear interaction. The Lyapunov exponents are given in
the the right panel of the Fig. \ref{multi} and have the values $(0.22685, 0.00431, -0.00431, -0.22685)$.
It may be noted that the highest Lyapunov exponent for $\Gamma=0$ is greater than the highest
Lyapunov exponent for $\Gamma=.01$ with all other conditions remaining the same. The numerical
results are described in detail in Ref. \cite{pkg-pr}.

The standard damped Duffing oscillator with a driving term is known to show chaotic
behaviour and a Hamiltonian formulation of the system is not known. However, for $H_D$,
there is neither any explicit damping term nor any driving term, yet the coupling between the
modes allows a chaotic regime. This may be explained as follows by analyzing the equations
of motion. In particular, the nonlinear terms in Eq. (\ref{duff-eqn}) with $\Gamma=0$ can not
be generated as the gradient of a potential function $V$, i.e. there is no solution to the coupled
partial differential equations,
\bea
\frac{\partial V}{\partial x}=-g x^3 , \ \frac{\partial V}{\partial y} =-3 g x^2 y
\eea
\noindent This implies that the system contains positional non-conservative forces\cite{marsden}
or curl-forces\cite{berry} which are known to admit chaotic behaviour. The driving is provided
by the coupling term $\beta_1 y$ in a non-standard way. This model may play a significant role
in the context of investigations on curl-forces admitting a Hamiltonian formalism
and quantum chaos, and further investigations in this direction are required.

\subsubsection{Landau Hamiltonian with balanced loss and gain:}

The coupled Duffing oscillator model in Eq. (\ref{v0-duff-eqn}) does not contain any velocity mediated coupling.
The observation on a possible relation between ${\cal{PT}}$ symmetry and existence of the equilibrium state
changes for system with velocity mediated coupling which may be explained by using Eq. (\ref{lan-equation}) which
describes Landau system with balanced loss and gain. The system is not ${\cal{PT}}$-symmetric. The second and the third terms in each
equation of (\ref{lan-equation}) breaks ${\cal{PT}}$ symmetry. However, the system admits periodic solutions in Region-I.
The observation is that if the linear part of equations of motion of a system contains velocity mediated interaction
and is non-${\cal{PT}}$-symmetric, the system may admit periodic solutions. It may noted that the system of equations
is invariant under the transformation:
\bea
x_1 \rightarrow x_2, x_2 \rightarrow - x_1, t \rightarrow - t, B \rightarrow - B.
\label{non-stand}
\eea
\noindent The transformation on the spatial co-ordinates correspond to a rotation by $\frac{\pi}{2}$ around
an axis perpendicular the `$x_1-x_2$'-plane, which can not be identified as discrete ${\cal{P}}$ transformation.
The transformation involving $t$ and $B$ correspond to time-reversal symmetry.

\subsubsection{Dimer \& Non-linear Schr$\ddot{o}$dinger Equation with balanced loss and gain:}

The following dimer model describes time-evolution of amplitudes of $x$ and $y$ in the
leading order of multiple time-scale analysis of Eq. (\ref{v0-duff-eqn}) for $\Gamma \ll 1, \beta
\ll 1$:\footnote{ Substitute $A \rightarrow \frac{1}{2} \Psi, T_2 \rightarrow t, \beta_0 \rightarrow 2 \beta,
\alpha \rightarrow \frac{8 \alpha}{3}$ in Eq. (43) of \cite{pkg-pr} to get this particular form of the
equation.}
\bea
i \frac{\partial \Psi}{\partial t} + \left ( i \Gamma_0 \sigma_3 + \beta \sigma_1 \right ) \Psi + \alpha
\begin{pmatrix} {\vert \psi_1 \vert}^2 \psi_1\\ 2 {\vert \psi_1 \vert}^2 \psi_2 + \psi_1^2 \psi_1^* \end{pmatrix}=0,
\ \ \Psi\equiv \begin{pmatrix} \psi_1\\ \psi_2 \end{pmatrix}
\label{dimer}
\eea
\noindent The time-reversal transformation is given by $t \rightarrow -t, i \rightarrow -i$, while ${\cal{P}}:
\Psi \rightarrow \sigma_1 \Psi$. The linear part of the equation, i.e. $\alpha=0$, is invariant under ${\cal{PT}}$
transformation. However, the system is non-${\cal{PT}}$-symmetric for $\alpha \neq 0$, yet it admits periodic solutions
in some region in the parameter-space\cite{pkg}. In particular, the Stokes variables,
\bea
Z_a= \frac{1}{2} \Psi^{\dagger} \sigma_a \Psi, \
R=\frac{1}{2} \Psi^{\dagger} \Psi=\sqrt{Z_1^2+Z_2^2+z_3^2}, \ \ a=1, 2, 3 
\label{stokes}
\eea
\noindent satisfy a solvable linear equations from which the time-periodic solutions $\Psi$ may be
constructed\cite{pkg-pr}.
Adding a dispersion term to Eq. (\ref{dimer}) leads to a coupled non-linear Schr$\ddot{o}$dinger equation
with balanced loss and gain,
\bea
i \frac{\partial \Psi}{\partial t} + \left ( i \Gamma_0 \sigma_3 + \beta \sigma_1 \right ) \Psi + 
\frac{\partial^2 \Psi}{\partial x^2} + \alpha
\begin{pmatrix} {\vert \psi_1 \vert}^2 \psi_1\\ 2 {\vert \psi_1 \vert}^2 \psi_2 + \psi_1^2 \psi_1^* \end{pmatrix}=0,
\label{nlse}
\eea
\noindent the linear part of which is ${\cal{PT}}$-symmetric, while the the nonlinear term explicitly breaks
${\cal{PT}}$-symmetry. The moments ${\cal{Z}}_a=\int dx \ Z_a, {\cal{R}}=\int R \ dx$ satisfy the same time-evolution
Eq. as satisfied by the Stokes variables given in Eq. (\ref{stokes})
for well behaved fields $\Psi$ vanishing at asymptotic infinity. Thus, the moments are time-periodic for
the same region for which the dimer model admits
periodic solutions. The spatial degree of freedom has been integrated out and the analytic expression for
$\Psi(x,t)$ can not be obtained from the time-dependence of the moments. However, the existence
of periodic solutions in some region of the parameter-space is ensured. 

\subsection{Non-${\cal{PT}}$-symmetric Non-Hamiltonian System}

\begin{figure}
\begin{minipage}{3in}
\centering
\includegraphics[width=3in]{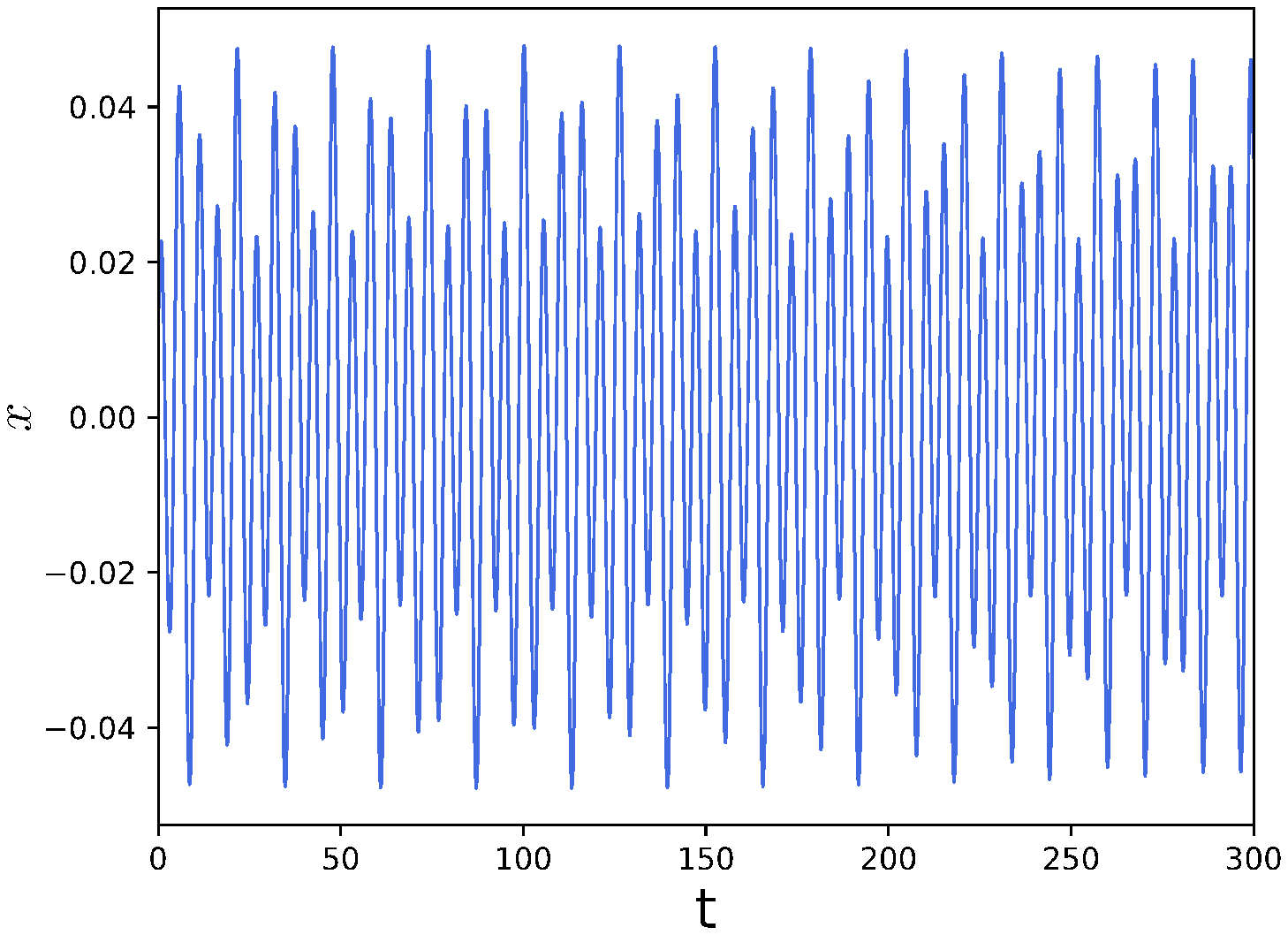}
\label{xtimeseries1}
\end{minipage} \hspace{.2in}%
\begin{minipage}{3in}
\centering
\includegraphics[width=3in]{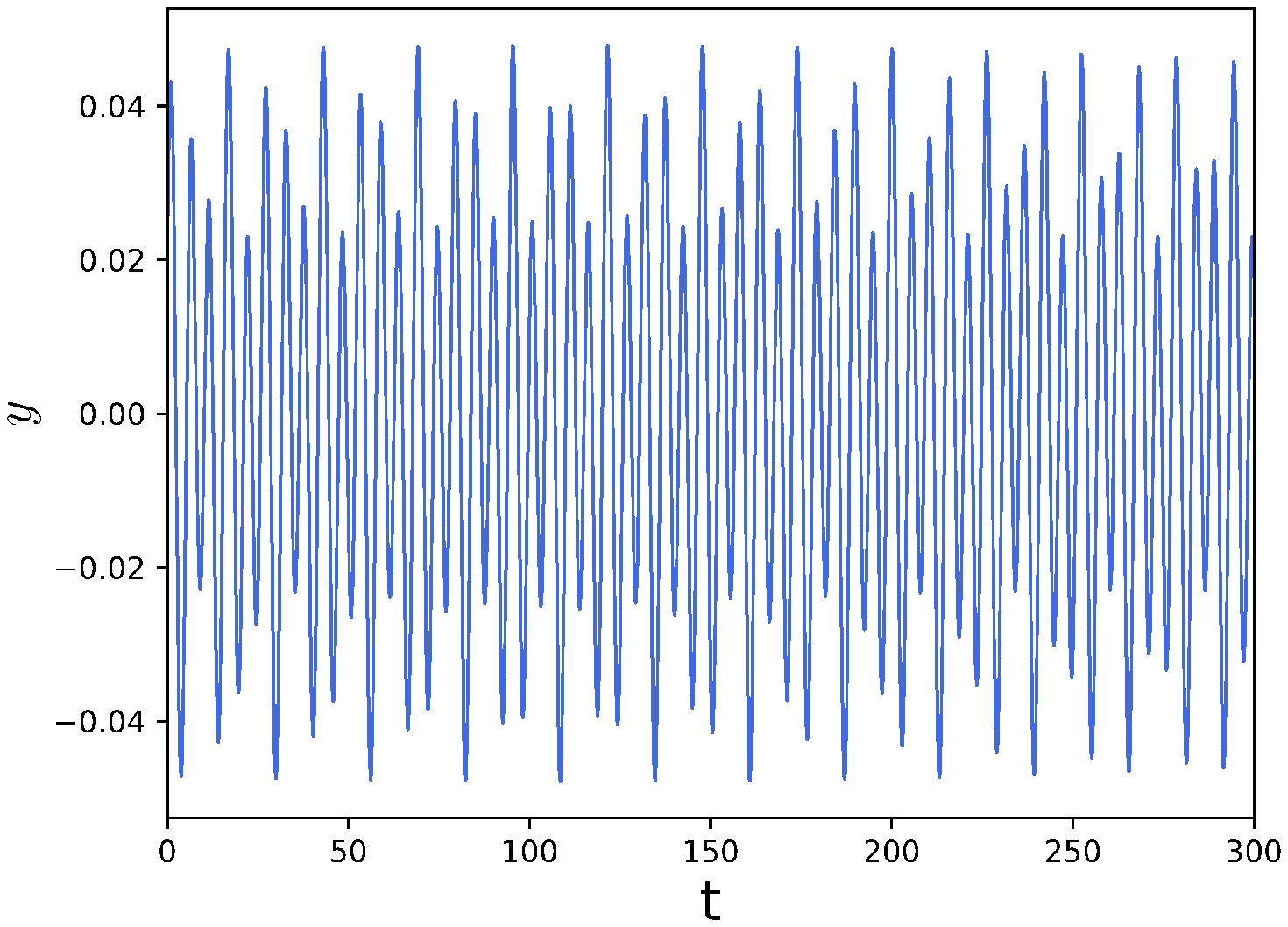}
\label{ytimeseries1}
\end{minipage}
\newline
\begin{minipage}{3in}
\centering
\includegraphics[width=3in]{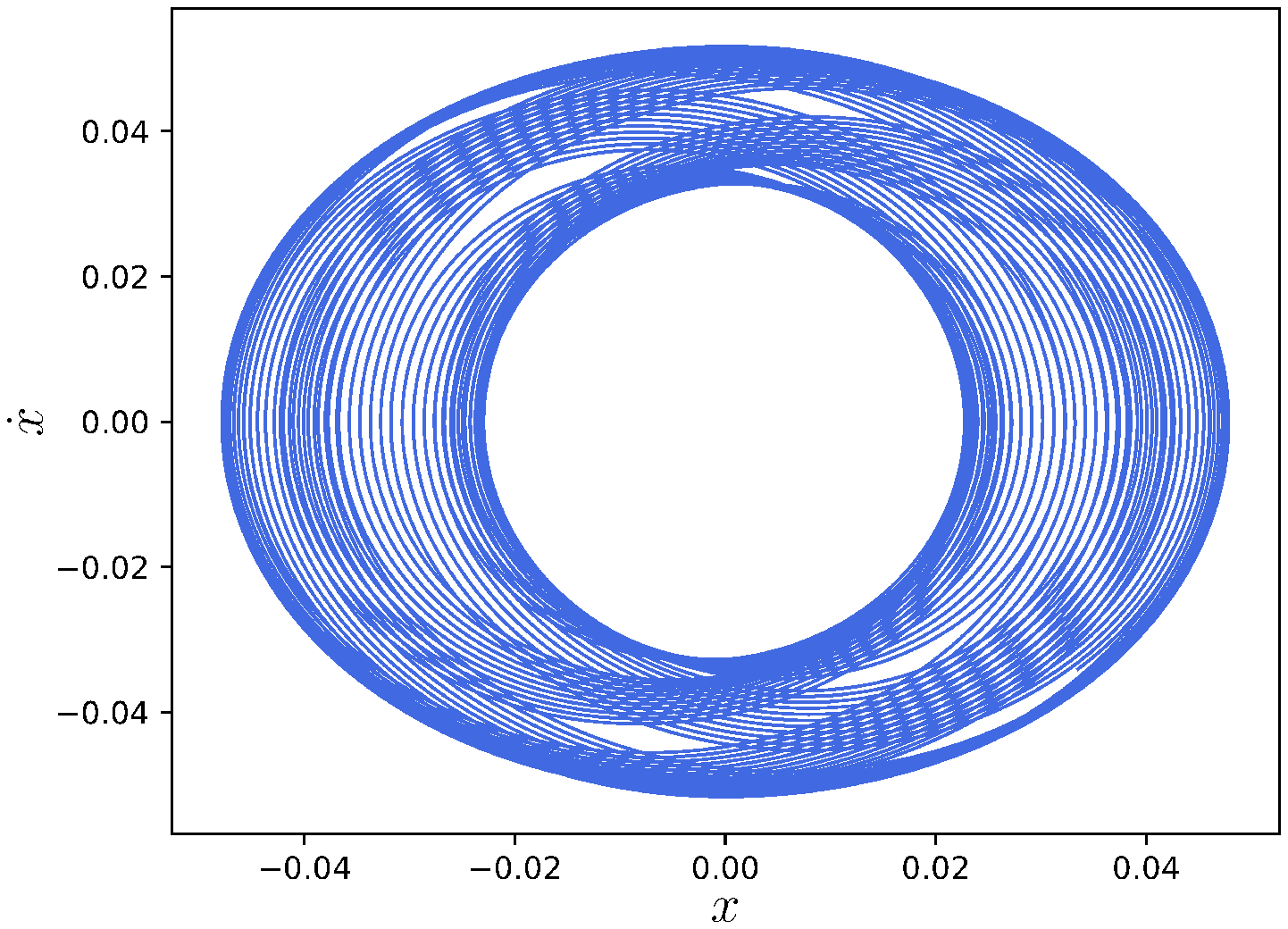}
\label{xtimeseries1}
\end{minipage} \hspace{.2in}%
\begin{minipage}{3in}
\centering
\includegraphics[width=3in]{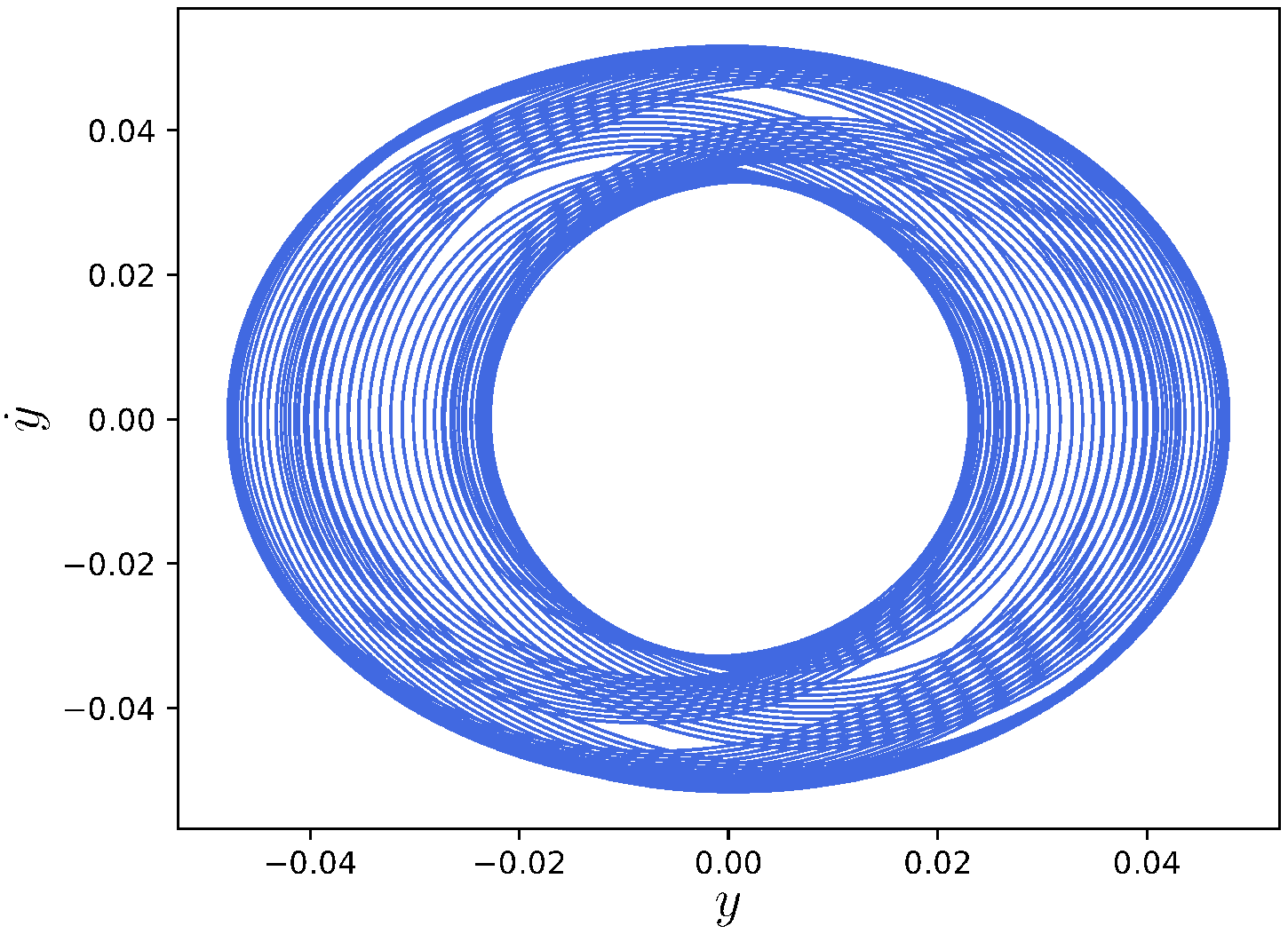}
\label{ytimeseries1}
\end{minipage}
\caption{(Color online)
Eq. (\ref{nh-do-eqn}) is solved with the initial conditions 
$x(0)=.01, y(0)=0.02, \dot{x}(0)=.03, \dot{y}(0)=.04$ for 
$\alpha_1=0.5, \alpha_2=1.0, \beta=0.5$ and $\Gamma=0.1$.
Time-series of $x$ and $y$ are plotted in the first row.
Plots of $\dot{x}$ vs. $x$ and $\dot{y}$ vs $y$ are given in the second row.}
\label{nh-do-fig}
\end{figure}

The central focus of this review is on Hamiltonian system. However, taking a detour from the main line
of discussions, an example of a non-${\cal{PT}}$-symmetric non-Hamiltonian System with balanced loss and
gain that admits periodic solutions is presented in this section. The system is described by the equations of motion,
\bea
&& \ddot{x} + 2\gamma\dot{x}+\omega^{2}x+\beta y + \alpha_1 x^{3}=0,\nonumber \\
&& \ddot{y} - 2\gamma\dot{y}+\omega^{2}y+\beta x + \alpha_2 y^3=0,
\label{nh-do-eqn}
\eea
\noindent which describe a damped Duffing oscillator linearly coupled to an anti-damped Duffing
oscillator with different nonlinear strengths. The gain and loss terms are equally balanced and the
flow in the position-velocity state space preserves its volume. However, no Hamiltonian for the
system is known for $\alpha_1 \neq 0$ and/or $\alpha_2 \neq 0$. The system is Hamiltonian for $\alpha_1=\alpha_2=0$
as well as ${\cal{PT}}$-symmetric\cite{cmb}.
The coupled Duffing oscillator model described by Eq. (\ref{nh-do-eqn}) is
${\cal{PT}}$-symmetric for $\alpha_1=\alpha_2$ and has been studied in detail in Ref.
\cite{cksk}. The system is non-Hamiltonian and admits periodic solutions in the unbroken ${\cal{PT}}$-regime.
The nonlinear term breaks ${\cal{PT}}$-symmetry for $\alpha_1 \neq \alpha_2$. It
appears that the solutions of Eq. (\ref{nh-do-eqn}) have not been investigated for
$\alpha_1 \neq \alpha_2$ for which the system is non-${\cal{PT}}$-symmetric.
The linear stability analysis predicts periodic solutions around the equilibrium
point $(0,0,0,0)$ in the position-velocity state space for
\bea
-\frac{1}{\sqrt{2}} < \Gamma < \frac{1}{\sqrt{2}}, \ \
4 \Gamma^2 \left ( 1 - \Gamma^2 \right ) < \beta^2 < 1\
\eea
\noindent This is confirmed by numerical analysis and time-series of $x$ and $y$ 
for $\alpha_1=0.5, \alpha_2=1.0, \beta=0.5, \Gamma=0.1$ and are shown in the first row of Fig. \ref{nh-do-fig}.
Plots of $\dot{x}$ vs $x$ and $\dot{y}$ vs. $y$ for the same values of the parameters are shown in the second
row of Fig. (\ref{nh-do-fig}). Numerical investigations show periodic solutions in a large region in the
parameter space. The periodic solution of non-Hamiltonian non-${\cal{PT}}$-symmetric coupled Duffing oscillator 
model is presented in this review for the first time and has not been discussed previously. 

\subsection{A Conjecture}

One important observation related to these investigations is that linear part of the equations for systems without
any velocity mediated coupling is necessarily ${\cal{PT}}$-symmetric in order to have periodic solutions, while the
nonlinear part may or may not be ${\cal{PT}}$-symmetric.
It may be conjectured at this point that {\it a non-${\cal{PT}}$ symmetric system with balanced loss-gain and without
any velocity mediated coupling may admit periodic solution if the linear part of the equations of motion is necessarily
${\cal{PT}}$ symmetric \textemdash the nonlinear interaction may or may not be
${\cal{PT}}$-symmetric. Further, systems with velocity mediated coupling among different
degrees of freedom need not be PT symmetric at all in order to admit periodic solutions.}
The conjecture has no contradiction with the known results on
${\cal{PT}}$-symmetric systems. There is no general result in a model independent
way to suggest that ${\cal{PT}}$-symmetry of a system is necessary in order to have
periodic solutions. On the contrary, it is known that non-Hamiltonian dimer model without
any ${\cal{PT}}$ symmetry due to imbalanced loss and gain admits stable nonlinear supermodes\cite{kominis}.
Similarly, within the mean field description of Bose-Einstein condensate, stationary ground state
is obtained for non-${\cal{PT}}$-symmetric confining potential\cite{lunt}. The results stated above 
and described in detail in Refs. \cite{pkg-pr,pr-pkg} are related to mechanical systems with finite degrees
of freedom, dimers and nonlinear Schr$\ddot{o}$dinger equation. Thus, the conjecture is supported by known
results from diverse areas of classical physics.

The conjecture is also consistent with ${\cal{PT}}$-symmetric quantum system due to the following reasons:
\begin{itemize}
\item Different realizations of the linear operator ${\cal{P}}$ and the anti-linear operator ${\cal{T}}$ are
allowed\cite{pkg-ds,haake} in quantum mechanics as long as ${\vert \langle \tilde{\phi}_i|\tilde{\psi}_i \rangle\vert}
={\vert \langle \phi| \psi \rangle \vert}$, where $\tilde{\phi}_1={\cal{P}} \phi, \tilde{\psi}_1= {\cal{P}} \psi,
\tilde{\phi}_2={\cal{T}} \phi, \tilde{\psi}_2= {\cal{T}} \psi$ and $\phi, \psi, \phi_i, \psi_i$ are state vectors
in the relevant Hilbert space. There may exist non-trivial realizations of ${\cal{P}}$ and ${\cal{T}}$ operators in
the corresponding quantum system such that it is ${\cal{PT}}$ invariant. An example of this non-standard
${\cal{T}}$-symmetry within the context of ${\cal{PT}}$-symmetric theory is discussed in
Ref. \cite{pkg-ds}.

\item The standard ${\cal{PT}}$-symmetry may be substituted with an appropriate anti-linear symmetry for which
consistent quantum description of a non-hermitian Hamiltonian admitting entirely real spectra and unitary
time-evolution is allowed. A pseudo-hermitian system\cite{am} admits anti-linear symmetry. Thus,
a non-${\cal{PT}}$-symmetric classical system with balanced loss and gain and admitting periodic solutions may,
upon quantization, become a pseudo-hermitian system.
\end{itemize}
\noindent There is no analogue of anti-linear symmetry and pseudo-hermiticy in the  classical physics. This
leads to fixing criterion for the existence of periodic solution of a system with balanced loss and gain solely
in terms of ${\cal{PT}}$ symmetry. However, any criterion based on ${\cal{PT}}$ symmetry is inadequate and incomplete.
A possible resolution of the problem may be to identify appropriate
${\cal{PT}}$-symmetry/anti-linear symmetry/pseudo-hermiticity of the corresponding quantized non-hermitian
Hamiltonian which may explain the entirely real spectra with unitary time-evolution. The second step is to take
the classical limit of the relevant anti-linear operator, which may not necessarily
reduce to the standard ${\cal{P}}$ and ${\cal{T}}$ transformation, and fix the criteria for the existence
of periodic solutions for the corresponding classical Hamiltonian.
It should be mentioned here that an implementation of the scheme is tricky and nontrivial, since there may be more than
one quantum system for a given classical Hamiltonian based on the quantization condition.
An intelligent resolution of the problem is desirable.

\section{Omitted topics}\label{ot}

The discussion so far is restricted to mechanical system with finite degrees
of freedom. The discrete and continuum models of non-linear
Schr$\ddot{o}$dinger equation with balanced loss and gain or its generalized versions
appear in diverse contexts and have been studied extensively in the literature
with interesting results\cite{rmp}. The nonlinear Dirac equation with balanced
loss-gain has also been investigated\cite{nde,nde1}.
A brief description on these topics is presented below.

\subsection{Non-linear Schr$\ddot{o}$dinger Equation with balanced loss and gain}

The non-linear Schr$\ddot{o}$dinger equation in one dimension is an integrable system with exact soliton
solutions\cite{mana,zs,ml-nlse,wang}. Several generalizations of non-linear Schr$\ddot{o}$dinger equation
have been considered with the growing interest and relevance on ${\cal{PT}}$
symmetric theory. The ${\cal{PT}}$ symmetry motivated generalizations of
non-linear Schr$\ddot{o}$dinger equation may be broadly classified into three major sub-areas:
(i) non-linear Schr$\ddot{o}$dinger equation with ${\cal{PT}}$-symmetric confining complex potential\cite{complexp}, 
(ii) non-local non-linear Schr$\ddot{o}$dinger equation\cite{abl,ds-nlse,ds-nlse-1} and (iii) non-linear Schr$\ddot{o}$dinger Equation
with balanced loss and gain\cite{rmp,dm,fa,igor-2,pd,alex,dias,meme}.
The topic non-linear Schr$\ddot{o}$dinger equation with balanced loss and gain is relevant for the present review for
which the literature is vast and the major results till 2016 are
described in Ref. \cite{rmp}. A recent result on a class of exactly solvable non-linear Schr$\ddot{o}$dinger equation with balanced loss and gain
is described below in brief.

The generic form of the non-linear Schr$\ddot{o}$dinger equation with balanced loss and gain is given by,
\bea
i \left ( I \frac{\partial}{\partial t} + i A \right ) \Psi +
\frac{\partial^2 \Psi}{\partial x^2} + V(x) \Psi +
G(\Psi, \Psi^*) \Psi=0, \ A:= B + i C,
\label{dimsum-1}
\eea
\noindent where $I$ is the $N \times N$ identity matrix, $\Psi$ is an $N$-component
complex scalar field and $G$ is an $N \times N$ hermitian matrix depending on the field $\Psi$. The
hermitian matrix $B$ describes linear coupling among different fields, while
the traceless hermitian matrix $C$ describes balanced loss and gain. The time-dependent gain-loss and linear
coupling terms may be considered by taking time-dependent matrices $C$ and $B$, respectively.
The matrix $G$ encodes nonlinear coupling among different fields and may be chosen depending
on the physical system. The external potential $V$ is generally taken as complex and ${\cal{PT}}$-symmetric.
However, the discussion in this article is restricted to real $V$. Various soliton solutions of
Eq. (\ref{dimsum-1}) for $V=0$ and specific forms of $G$ have been obtained under
certain reductions\cite{rmp,dm,fa,igor-2,pd,alex,dias}. Recently, exactly
solvable models of  non-linear Schr$\ddot{o}$dinger equation  with balanced loss and gain have been constructed in Ref. \cite{meme}
for $G=I \ \Psi^{\dagger} M \Psi$, where $M$ is a hermitian matrix
independent of fields. The space-time modulation of the cubic nonlinearity
may be introduced via the matrix $M$.

The continuity equation contains source and sink terms proportional to $\Psi^{\dagger} C \Psi$:
\bea
\frac{\partial \tilde{\rho}}{\partial t} + \frac{\partial \tilde{J}}{\partial x}= 2 \Psi^{\dagger} C \Psi, \ \
\tilde{\rho}=\Psi^{\dagger} \Psi, \ \ \tilde{J}= - i \left [ \Psi^{\dagger} \frac{\partial \Psi}{\partial x} -
\frac{\partial \Psi^{\dagger}}{\partial x} \Psi \right ]
\eea
\noindent In general, the total density $\tilde{Q}=\int dx \tilde{\rho}$ is not conserved  
even for well-behaved $\Psi$ vanishing at asymptotic infinity, $\frac{\partial \tilde{Q}}{\partial t}
=2 \int dx \Psi^{\dagger} C \Psi$. There may be specific field configurations with additional properties,
like $\Psi^{\dagger} C \Psi$ being an odd function of $x$,  for which $\tilde{Q}$ is conserved. It is
important to note that a non-standard continuity equation with the associated conserved quantity can be
derived for the special case of an $\eta$-pseudo-hermitian\cite{am} $A$, i.e. $A^{\dagger} = \eta A
\eta^{-1}$, 
\bea
\frac{\partial \rho}{\partial t} + \frac{\partial J}{\partial x}=0,
\ \ \rho = \Psi^{\dagger} \eta \Psi, \ J= - i \left [ \Psi^{\dagger} \eta \frac{\partial \Psi}{\partial x}
 - \frac{\partial \Psi^{\dagger}}{\partial x} \eta \Psi \right ].
\eea
\noindent The quantity $Q=\int dx \rho$ is a conserved quantity for the well-behaved fields $\Psi$
vanishing at infinity. The continuity equation holds irrespective of whether the matrix $\eta$ is 
positive-definite or indefinite. A positive-definite $\eta$ ensures a positive-definite $Q$. Further,
it is known\cite{am} that a pseudo-hermitian matrix $A$ with respect to a positive-definite $\eta$
admits entirely real eigenvalues which is required for the existence time-periodic solution of $\Psi$.
It may be noted that Eq. (\ref{dimsum-1}) may or may not depend on $\eta$ and
if it depends on $\eta$ at all, it should be through the matrix $G$.

The system described by Eq. (\ref{dimsum-1}) admits a Lagrangian $L=\int dx {\cal{L}}$,
\bea
{\cal{L}}_F  = 
\frac{i}{2} \left [ \Psi^{\dagger} M \left ( D_0 \Psi\right ) - 
(D_o \Psi)^{\dagger} M \Psi \right ]
 -  \frac{\partial \Psi^{\dagger}}{\partial x}
M\frac{\partial \Psi}{\partial x} + V(x) \Psi^{\dagger} M \Psi
+ W(\Psi, \Psi^{\dagger}) +\Psi^{\dagger} F_1 \Psi,
\label{lag-nlse}
\eea
\noindent where the matrix $G(\Psi,\Psi^{\dagger})$ and the potential $W(\Psi, \Psi^{\dagger})$ are
related by the equation,
\bea
G= M^{-1} \frac{\partial W}{\partial \Psi^{\dagger}}
\label{gwrel}
\eea
\noindent and the hermitian matrix $M$ is independent of field $\Psi$.
The operator $D_0:=I \frac{\partial}{\partial t} + i A $ has formal
resemblance with the temporal component of covariant derivative with non-hermitian gauge potential
$A$ and the anti-hermitian matrix $F_1:=\frac{1}{2} \left ( A^{\dagger}M-M A \right )$.
There may be alternative Lagrangian formulations of Eq. (\ref{dimsum-1}) for specific $G$. This
particular formulation is chosen due to its conceptual similarity with ${\cal{L}}$ of
Eq. (\ref{lag}) describing mechanical system with finite degrees of freedom \textemdash presence
of fictitious gauge potential and a metric is common to both the formulations.
The conjugate momenta corresponding to $\Psi$ and $\Psi^{\dagger}$ are $\Pi_{\Psi}=
\frac{i}{2} \Psi^{\dagger} M$ and $\Pi_{\Psi^{\dagger}}=- \frac{i}{2} M \Psi$, respectively.
The Hamiltonian density ${\cal{H}}_F$ corresponding to ${\cal{L}}_F$ has the form,
\bea 
{\cal{H}}_F = \frac{\partial \Psi^{\dagger}}{\partial x} M \frac{\partial \Psi}{\partial x}
-V(x) \Psi^{\dagger} M \Psi - W(\Psi,\Psi^{\dagger}) + \Psi^{\dagger} M A \Psi
\label{hami-nlse}
\eea
\noindent In general, the Hamiltonian ${\cal{H}}_F$ is  complex-valued and the quantized Hamiltonian
is expected to be non-hermitian. However, for the case of an $M$-pseudo-hermitian $A$, ${\cal{H}}_F$ is
real-valued and the corresponding quantum Hamiltonian is hermitian with suitable quantization condition.
An $M$-pseudo-hermitian matrix $A$ may be used to define an $M$-pseudo-unitary\cite{ali} matrix $U:=e^{-i A}$,
\bea
U^{\dagger} M U =  M  \Leftrightarrow A^{\dagger}=M A M^{-1}
\label{e75}
\eea
\noindent The Hamiltonian ${\cal{H}}_F$ is invariant under pseudo-unitary transformation
$\Psi \rightarrow U \Psi$ provided $W$ is invariant under this transformation. For example,
the Hamiltonian is invariant under pseudo-unitary transformation for the choice
$W \equiv W(\Psi^{\dagger} M \Psi)$. It should be noted that two systems connected by a unitary
transformation can be considered as gauge equivalent. However, the same is not true for systems
connected by pseudo-unitary transformations \textemdash the physical observable like density,
square of the width of the wave-packet and its speed of growth have different expressions.
The pseudo-unitary transformation can be used to construct solvable models\cite{meme}.

\subsubsection{${ V=0}$ and ${ W= \frac{\delta}{2} \ \left (\Psi^{\dagger} M \Psi \right )^2}$:}

The  non-linear Schr$\ddot{o}$dinger equation  in Eq. (\ref{dimsum-1}) takes the form,
\bea
i \left ( I \frac{\partial}{\partial t} + i A \right ) \Psi +
\frac{\partial^2 \Psi}{\partial x^2} + \delta \left ( \Psi^{\dagger} M  \Psi \right ) \Psi=0,
\label{dimsum-2}
\eea
\noindent Defining $\Psi=e^{-i A} \Phi$ and using Eq. (\ref{e75}), Eq. (\ref{dimsum-1}) is mapped\cite{meme}
to the equation,
\bea
i \Phi_t =  - \Phi_{xx} -\delta \left ( \Phi^{\dagger} M \Phi \right ) \Phi.
\label{mana-eq}
\eea
\noindent Eq. (\ref{mana-eq}) is exactly solvable and with appropriate unitary transformation
followed by a scaling\cite{meme}, it can be brought to the canonical form of
Manakov-Zakharov-Schulman system\cite{mana,zs}. The pseudo-unitary mapping can be used to find
exact solutions of Eq. (\ref{dimsum-1}) with $V=0$ and $G=\Psi^{\dagger} M \Psi$. Exactly solvable models with
analytic expression for power-oscillation have been constructed in Ref. \cite{meme}
by using this mapping. If the non-hermitian matrix $A$ is not pseudo-hermitian, then 
the mapping can be used to construct exact solutions for a non-autonomous system\cite{meme}.
Further, the mapping can be used to construct a variety of solvable models with
interesting physical properties\cite{sg-pkg} like time-dependent gain-loss terms, space-time
modulated nonlinear interaction and external confining potential\cite{sg-pkg}.

\subsection{Oligomer with balanced loss and gain}

The discrete  non-linear Schr$\ddot{o}$dinger equation  with finite number of lattice points $N$ or similar equation
with generalized interaction term and self-trapping phenomena is popularly known
as oligomer and the $N=2$ is termed dimer. Such equations arise in the
study of ${\cal{PT}}$-symmetric lattices\cite{rmp}, nonlinear dynamics of
molecules\cite{scott,susanto}, optics\cite{delfino}, Landau-Lifshitz equation\cite{lan-lif} etc..
The time-evolution
of amplitude of waves arising in a mechanical and/or extended system under
various approximation schemes may also be modelled as oligomer. For example, 
a multiple time-scale analysis of Eq. (\ref{v0-duff-eqn}) leads to
the dimer model (\ref{dimer}) describing the time-evolution of amplitudes
in the leading order of the perturbation. The same recipe for studying
the time-evolution of amplitudes of nonlinear oscillators has been used
earlier in Refs. \cite{igor,khare}. 

The equation satisfied by an oligomer has the general form,
\bea
i \left ( I \partial_t + i A \right ) \Psi + G(\Psi, \Psi^*) \Psi=0
\label{dimsum}
\eea
\noindent It may be noted that Eq. (\ref{dimsum-1}) for vanishing external potential $(V=0)$ and
no dispersion term $\frac{\partial^2 \Psi}{\partial x^2}$ reduces to the oligomer model in Eq. (\ref{dimsum}).
The matrices $B$, $C$ and $G(\Psi, \Psi^*)$ have the same interpretation as in the case of dimer model.
The discrete  non-linear Schr$\ddot{o}$dinger equation with balanced loss and gain is reproduced for a tridiagonal $B$ with
$[G]_{ij}= \alpha \delta_{ij} {\vert \Psi \vert}^2$, where the real parameter
$\alpha$ denotes the strength of the nonlinear interaction. There are specific
forms of $B$, $C$ and $G$ for which a Hamiltonian description is possible
and the system may even be integrable\cite{igor,khare,igor-1,zezyulin}. The
literature on dimer with balanced loss and gain is vast and results with physical settings are
nicely reviewed in Ref. \cite{rmp}. A new class of exactly solvable dimer models
which has not been discussed earlier is presented below.

A generic Lagrangian-Hamiltonian formulation of Eq. (\ref{dimsum}) may be presented by
continuing with the general development for mechanical system with finite degrees of freedom
and for the extended system. In particular, the Lagrangian and Hamiltonian are,
\bea
&& {\cal{L}}_O  = \frac{i}{2} \left [ \Psi^{\dagger} M \left ( D_0 \Psi\right ) - 
(D_o \Psi)^{\dagger} M \Psi \right ]
+ W(\Psi, \Psi^{\dagger}) +\Psi^{\dagger} F_1 \Psi,\nonumber \\
&& {\cal{H}}_O =  - W(\Psi,\Psi^{\dagger}) + \Psi^{\dagger} M A \Psi\nonumber
\label{lag-hami-oligomer}
\eea
\noindent where $G$ and $W$ are related via Eq. (\ref{gwrel}). The conjugate momenta corresponding to
$\Psi$ and $\Psi^{\dagger}$ are $\Pi_{\Psi}= \frac{i}{2} \Psi^{\dagger} M$ and $\Pi_{\Psi^{\dagger}}=
- \frac{i}{2} M \Psi$, respectively. A new class of solvable models are obtained
for the choice of $W=\frac{\delta}{n+1} \left (\Psi^{\dagger} M \psi \right )^{n+1}$ for which Eq. (\ref{dimsum}) takes the form,
\bea
i \left ( I \partial_t + i A \right ) \Psi + \delta \left ( \Psi^{\dagger} M \Psi \right )^n \Psi=0
\label{dimsum-olig}
\eea
\noindent and is exactly solvable for a $M$-pseudo-hermitian $A$. In particular, substituting
$\Psi=e^{-i A t} \Phi$ in eq. (\ref{dimsum-olig}), $\Phi$ satisfies the equation,
\bea
i \partial_t \Phi + \delta (\Phi^{\dagger} M \Phi)^n \Phi=0.
\eea
\noindent It immediately follows that $\Phi^{\dagger} M \Phi$ is a conserved quantity and denoting
its value at $t=0$ as the real constant $C$, i.e. $\Phi^{\dagger}(0) M \Phi(0)=C$, $\Phi$ is solved as,
\bea
\Phi(t)=W e^{i \delta C^n t},  
\eea 
\noindent where $W$ is an arbitrary $N$-component constant complex vector. For the simplest case of a dimer,
i.e. $N=2$, an example of $M$-pseudo-hermitian $A$ may be presented as,
\bea
A=\beta^* \sigma_+ + \beta \sigma_- + i \Gamma \sigma_3, \
M= \alpha_0 I_2 + \alpha^* \sigma_+ + \alpha \sigma_-
\eea
\noindent where $I_2$ is the $2 \times 2$ identity matrix and $\sigma_{\pm}=\frac{1}{2} \left (\sigma_1
\pm \sigma_2 \right )$. The condition for a $M$-pseudo-hermitian $A$
with positive definite $M$ has been derived in Ref. \cite{meme} in terms of the complex parameters
$\beta={\vert \beta \vert} e^{i \theta_{\beta}}, \alpha={\vert \alpha \vert}e^{i \theta_{\alpha}}$ and
the real parameters $\Gamma, \alpha_0$ as,
\bea
\frac{\alpha_0}{\vert \alpha \vert} = \frac{\vert \beta \vert}{\Gamma} \sin(\theta_{\alpha}-\theta_{\beta})
> 0.
\eea
\noindent Using the expression for $U:=e^{-i A t}$ in Ref. \cite{meme}, the solutions for $\Psi$ are obtained
as,
\bea
&& \Psi_1= e^{i \delta C^n t} \left [ W_1 \left ( \cos(\theta t) +\frac{\Gamma}{\theta} \sin (\theta t)
\right ) -\frac{i W_2 \beta^*}{\theta} \sin (\theta t) \right ]\nonumber \\
&& \Psi_2= e^{i \delta C^n t} \left [ -\frac{i W_1 \beta}{\theta} \sin (\theta t)
+ W_2 \left ( \cos(\theta t) - \frac{\Gamma}{\theta} \sin (\theta t) \right ) \right ]
\eea
\noindent where $\theta=\sqrt{{\vert \beta \vert}^2-\Gamma^2}$. The solutions are periodic for
${\vert \beta \vert} > \Gamma$ and unbounded for ${\vert \beta \vert} \leq \Gamma$. The matrix
$U$ is unitary for $\Gamma=0$ and $\Psi^{\dagger} \Psi$ is independent of time. However, for
$\Gamma \neq 0$, $\Psi^{\dagger} \Psi$ is periodic in time. Results for higher values of $N$
may be obtained in a similar way. 

\subsection{Nonlinear Dirac Equation with balanced loss and gain}

The nonlinear Dirac equation appears in diverse branches of modern science\cite{soler,gn,thirring,nde0,par}, although it has not
been studied as much as its non-relativistic counterpart, namely non-linear Schr$\ddot{o}$dinger equation. With the
advent of ${\cal{PT}}$-symmetric theory, several nonlinear Dirac equations with balanced loss and gain have been considered 
in the literature\cite{nde,nde1}. It may be recalled that the balanced loss and gain terms for the case of  non-linear Schr$\ddot{o}$dinger equation  is
introduced via a non-hermitian mass term in the Hamiltonian. The same approach is taken for the construction
of nonlinear Dirac equation with balanced loss and gain. The Lagrangian density for the Dirac equation has the generic form
\bea
{\cal{L}}_D= \bar{\Psi} \left ( i \gamma^{\mu} \partial_{\mu} - m_1 - m_2 \gamma_5 \right ) \Psi
- F(\bar{\Psi}, \Psi) 
\eea
\noindent with the non-hermitian mass term $\bar{\Psi} \left (m_1 + m_2 \gamma_5 \right )\Psi$\cite{nde3}
and Lorentz invariant nonlinear interaction term $F$. The interaction term $F$ may be chosen depending
on the physical settings. For example,  $F_1 \equiv F_1(\bar{\Psi} \Psi)$ reproduces Soler model\cite{soler}
for $m_2=0$ and in $1+1$ space-time it is also known as Gross-Neveu\cite{gn} model. The nonlinear interaction
$F_2\equiv F_2(\bar{\Psi} \gamma_5 \Psi)$ depends on pseudo-scalar $\bar{\Psi} \gamma_5 \Psi$. Similarly, the
choice $F=J_{\mu} J^{\mu}$ with the current $J_{\mu}=\bar{\Psi}\gamma_{\mu} \Psi$ introduces vector-type
self-interaction and the system describes
massive Thirring Model\cite{thirring} for $m_2=0$. The interaction term $F$ need not be Lorentz invariant for
non-relativistic systems with equations having formal resemblance with the Dirac equation.
The Hamiltonian is complex-valued for $m_2 \neq 0$,
\bea
{\cal{H}}_D=\bar{\Psi} \left ( i \vec{\gamma} \cdot \vec{\nabla} + m_1 + m_2 \gamma_5 \right ) \Psi +
F(\bar{\Psi}, \Psi),
\eea
\noindent and the quantum Hamiltonian is non-hermitian irrespective of whether $F$ is hermitian or not.

The equation of motion following from the Lagrangian reads,
\bea
\left ( i \gamma^{\mu} \partial_{\mu} - m_1 - m_2 \gamma_5 \right ) \Psi
- \frac{\partial F}{\partial \bar{\Psi}}=0.
\label{3d-deqn}
\eea
\noindent The balanced loss-gain terms can be seen explicitly
with the following representation of the $\gamma$ matrices,
\bea
\gamma^0= \sigma_1 \otimes I_2, \ \gamma^j= i \sigma_3 \otimes \sigma_j, \ \
\gamma_5:=i \gamma^0 \gamma^1 \gamma^2 \gamma^3= \sigma_2 \otimes I_2,  
\label{gamma-rep}
\eea
\noindent for which Eq. (\ref{3d-deqn}) reads,
\bea
i \left ( I_4 \frac{\partial}{\partial t} + i {\cal{M}} \right ) \Psi
= - i \sigma_2 \otimes \vec{\sigma} \cdot \vec{\nabla} \Psi + \gamma^0 \frac{\partial F}{\partial \bar{\Psi}},
\ \
{\cal{M}}:= \left ( m_1 \sigma_1 + i m_2 \sigma_3 \right ) \otimes I_2
\eea
\noindent The term appearing in the matrix ${\cal{M}}$ with the co-efficient $m_1$ describes linear coupling
among different components of the spinor $\Psi$, while the term with the co-efficient $m_2$ describes
balanced loss and gain. The loss-gain terms may be hidden by using a unitary equivalent representations
of the $\gamma$ matrices. For example, the loss-gain terms are hidden in the Dirac-Pauli representation.
The mass-matrix ${\cal{M}}$ is non-hermitian
with doubly degenerate eigenvalues $\lambda_{\pm}=\pm \sqrt{m_1^2-m_2^2}$ which are real provided
${\vert m_2 \vert} < {\vert m_1 \vert}$. The mass-gap $2 \sqrt{m_1^2-m_2^2}$ vanishes for $m_2=\pm m_1$,
signalling the existence of zero mode for the linear problem, i.e. $F=0$. The ansatz
$\Psi = W e^{i(\vec{k} \cdot \vec{r}  - \omega t)}$ with $F=0$ gives the dispersion $\omega^2=k^2 +
m_1^2-m_2^2$, where $W$ is a four-component constant complex vector. The frequency $\omega$ is real for
real mass, i.e. ${\vert m_2 \vert} < {\vert m_1 \vert}$ and $\Psi$ is periodic in time.

The charge $\tilde{Q}=\int d^3x \Psi^{\dagger} \Psi$ is not a conserved quantity due to the non-hermitian
mass term. However, a non-standard conserved charge $Q=\int d^3x \Psi^{\dagger} \eta \Psi$ can be introduced if
a hermitian matrix $\eta$ exists satisfying the following conditions:
\bea
\left [ \eta, \gamma^0 \gamma^i \right ] = 0 \ \forall \ i, \ {\cal{M}}^{\dagger}=\eta {\cal{M}} \eta^{-1}, \ \
W^{\dagger}=W, \ W \equiv \Psi^{\dagger} \eta \gamma^0 \frac{\partial F}{\partial \bar{\Psi}}
\eea 
\noindent The ansatz $\eta= \delta_1 I_4 + \delta_2 \sigma_2 \otimes I_2$ solves the first equation for arbitrary
real constants $\delta_1$ and $\delta_2$, since $\gamma^0 \gamma^i= \sigma_2 \otimes \sigma_i$. The second equation
demands that the mass-matrix ${\cal{M}}$ is $\eta$-pseudo-hermitian, which fixes $\delta_1=1,
\delta_2=\frac{m_2}{m_1}$ so that $\eta=I_4 + \frac{m_2}{m_1} \gamma_5$. It may be noted that $\eta$ is
positive-definite for ${\vert m_1 \vert} > {\vert m_2 \vert}$ and is singular for $m_2=\pm m_1$.
The conserved current $J^{\mu}$ has the expression $J^{\mu}= \bar{\Psi} \gamma^{\mu} \eta \Psi$
leading to the conservation of $Q$ for $F=0$. This relation for the linear Dirac equation has been obtained
earlier\cite{alex-1}. The probability density $J^0$ receives contribution only from right- or left-handed degrees
of freedom for $m_2=m_1$ and $m_2=-m_1$, respectively\cite{alex}. The hermiticity of $W$ has to be ensured such
that $J^{\mu}$ can also be taken as the conserved current for the nonlinear Dirac equation, i.e. $F \neq 0$. There
are choices of $F$ for which $W$ is hermitian. For example, $W$ is hermitian for $F \equiv F(\bar{\Psi} M \Psi)$
provided the constant matrix $M$ satisfies the condition,
\bea
\left [ M, \gamma^0 \right ] + \frac{m_2}{m_1} \left \{ M, \gamma^0 \gamma_5 \right \} =0.
\eea
\noindent  One important solution of the above equation is $M_1 =\eta$. The nonlinear interaction
$F$ contains both Lorentz scalar and pseudo-scalar interactions, since $\bar{\Psi} \eta \Psi
=\bar{\Psi} \Psi + \frac{m_2}{m_1} \bar{\Psi} \gamma_5 \Psi$. A more general solution can be
constructed by taking $M_2=\sigma_1 \otimes m$, where $m$ is an arbitrary $2 \times 2$ hermitian matrix.
It should be emphasized here that the existence of a conserved charge with its interpretation as the probability
density is not necessary in the study of nonlinear Dirac equation, since one is dealing with relativistic
field theory instead of quantum mechanics. The existence of a conserved charge corresponds to an internal symmetry
of the system. However, an appropriate number operator should be defined in the corresponding quantum field theory.
The investigations on nonlinear Dirac equation with balanced loss and gain are mainly restricted to its
classical solutions and stability in $1+1$ dimensions. Several choices of $F$ maintaining Lorentz invariance
and ${\cal{PT}}$-symmetry have been considered in Refs. \cite{nde,nde1} which admit stable periodic as well
as soliton solutions. 

\section{Summary \& Discussions}\label{sd}

Classical Hamiltonian systems with balanced loss and gain have been reviewed in this article. The emphasis is on
mechanical system with finite degrees of freedom. The criteria for a mechanical system to be identified as
a system with balanced loss and gain, irrespective of whether a Lagrangian-Hamiltonian formulation is admissible or not, has been
presented in terms of the volume conservation of flow in the position-velocity state-space. The Hamiltonian
formulation for systems with space-dependent balanced loss and gain has been introduced for arbitrary number of particles and
generic potential. It has been shown that the loss-gain terms may be removed completely through appropriate
co-ordinate transformations with its effect manifested in modifying the strength of the velocity-mediated
coupling. This mapping is inherent to the generic Hamiltonian system with balanced loss and gain and does not depend on any
specific form of the potential or the number of particles. In general, the quadratic term in momenta in
the Hamiltonian is not positive-definite leading to instabilities. The effect of the Lorentz interaction
in improving the stability of classical solutions as well as allowing a possibility of defining the
corresponding quantum problem consistently on the real line, instead of within Stokes wedges, has also been
discussed.

The system with $N=2m$ particles admits at least $m+1$ integrals of motion for a potential having specified
type of translational or rotational symmetry, thereby implying that the system
is at least partially integrable for $N >2$ and completely integrable for $N=2$. Several exactly solvable models
based on translational and rotational symmetry and specific form of potentials have been discussed which include
coupled cubic oscillators, Landau Hamiltonian etc. The Lorentz
interaction appear naturally in the Landau Hamiltonian and there are regions in the parameter
space where the quadratic term in momenta in the Hamiltonian is positive-definite. This is also the region
for which stable classical solutions are obtained. The solution is same as the standard Landau problem with
a reduced cyclotron frequency due to the loss-gain term.

An example of Hamiltonian chaos within the framework of a model of coupled Duffing oscillator with balanced loss and gain has
been discussed. The dynamical properties of the system are rich \textemdash three out of five equilibrium
points are stable and admits periodic solutions around these equilibrium points. The damped undriven Duffing
oscillator is linearly coupled to another anti-damped oscillator with variable angular frequency depending
on the degree of freedom of the Duffing oscillator. This coupling acts effectively as a driving term, albeit
in a nontrivial way. The chaotic behaviour is seen beyond a critical value of this coupling strength.
The chaotic behaviour persists even for vanishing gain-loss terms, thereby providing an example of Hamiltonian
chaos for coupled Duffing oscillator without any explicit damping and driving terms. It has been argued
that the system contains positional non-conservative force\cite{marsden} or the curl-force\cite{berry}, thereby
compensating the effect of damping even if $\gamma=0$. The coupling to the oscillator with variable frequency
provides the effect of driving term. The corresponding quantum system may provide some insight into quantum chaos.

The role of ${\cal{PT}}$-symmetry on the existence of periodic solution in systems with balanced loss and gain
has been critically
analyzed. Examples from many-particle mechanical systems, dimer models and nonlinear Schr$\ddot{o}$dinger equations
without any ${\cal{PT}}$-symmetry is analyzed with the understanding that ${\cal{P}}$ corresponds to
linear transformation only. This is because the Lorentz transformation is linear and no signature of its
violation has been seen in nature. Moreover, ${\cal{PT}}$-symmetric quantum mechanics dwells on ${\cal{CPT}}$
norm, which at a more fundamental level is expected to correspond to the ${\cal{CPT}}$ invariance
of a local hermitian Lorentz invariant theory. Based on the observations, it has been conjectured that
non-${\cal{PT}}$-symmetric system  with balanced loss-gain and without any velocity mediated interaction
may admit periodic solution if the linear part of the equations is necessarily ${\cal{PT}}$ symmetric \textemdash
the nonlinear interaction may or may not be ${\cal{PT}}$-symmetric.Further,
systems with velocity mediated coupling among different degrees of freedom need not be ${\cal{PT}}$ symmetric at
all in order to admit periodic solutions. This conjecture has no contradiction with the
formulation of ${\cal{PT}}$-symmetric quantum mechanics \textemdash the corresponding quantum system may be
pseudo-hermitian or invariant under generalized time-reversal symmetry which has no analog in classical
mechanics. The criteria for the existence of periodic solutions in terms of standard ${\cal{PT}}$-symmetry
alone is not sufficient. Investigations in this direction is desirable so that a large number of non-${\cal{PT}}$
symmetric system may be included in the mainstream of investigations with possible technological applications.

The central focus of this review is on mechanical system with finite degrees of freedom. However,
there is significant advancement in the fields of oligomers, nonlinear Schr$\ddot{o}$dinger and Dirac equations
with balanced loss and gain. The developments in the context of oligomers and nonlinear Schr$\ddot{o}$dinger
equations are summarized recently in Ref. \cite{rmp} and excluded for extensive discussions in this review.
A very recent result on  non-linear Schr$\ddot{o}$dinger equation  with balanced loss and gain related to the construction of exactly solvable models
via non-unitary transformation has been discussed. This mapping can be used to construct a variety of solvable models with
interesting physical properties\cite{sg-pkg}. The same technique is used to construct an exactly solvable dimer with
balanced loss and gain which has not appeared earlier in the literature. Results related to nonlinear Dirac equations 
with balanced loss and gain are mentioned briefly.


\section{Acknowledgments}

This work is supported by a grant ({\bf SERB Ref. No. MTR/2018/001036})
from the Science \& Engineering Research Board(SERB), Department of Science
\& Technology, Govt. of India under the {\bf MATRICS} scheme. The Author
would like to thank Debdeep Sinha, Puspendu Roy and Supriyo Ghosh for discussions
and collaboration on the topic.


\end{document}